\documentclass[reprint,showpacs,groupedaddress,eqsecnum,amsfonts,amsmath,amssymb,aps,prd]{revtex4-1}
\usepackage{graphicx}
\usepackage{latexsym}
\usepackage{amssymb}
\usepackage{mathrsfs}
\usepackage{amsmath}
\usepackage{amsthm}
\usepackage{color}
\usepackage{dcolumn}% Align table columns on decimal point
\usepackage{bm}% bold math
\newcommand{\omits}[1]{}
\definecolor{dyellow}{rgb}{1.,0.8,.0}
\definecolor{myblue}{rgb}{.1,.1,.7}
\definecolor{dcyan}{rgb}{.0,.6,.6}
\definecolor{dmagenta}{rgb}{0.6,0.0,0.6}
\definecolor{brown}{rgb}{0.6,0.2,0.}
\definecolor{darkblue}{rgb}{.0,.0,0.5}
\definecolor{darkred}{rgb}{0.75,0.0,0.0}
\definecolor{orange}{rgb}{1.,.6,.0}
\definecolor{dorange}{rgb}{0.8,.4,.0}
\definecolor{darkgreen}{rgb}{0.0,0.6,0.0}
\definecolor{purple}{rgb}{.4,.0,.4}
\definecolor{grey}{rgb}{0.5,0.5,0.5}
\def\delete{\color{grey}}
\begin{document}
\hyphenpenalty=1000
%\preprint{APS/123-QED}
\title{Multipole analysis for linearized $f(R)$ gravity with irreducible Cartesian tensors}

\newcommand*{\PKU}{Institute of High Energy Physics and Theoretical Physics Center for Science Facilities,
Chinese Academy of Sciences, Beijing, 100049, People's Republic of China}\affiliation{\PKU}
\newcommand*{\INFN}{INFN, Sez. di Pavia, via Bassi 6, 27100 Pavia, Italy}\affiliation{\INFN}
\newcommand*{\CICQM}{}\affiliation{\CICQM}
\newcommand*{\CHEP}{}\affiliation{\CHEP}

\author{Bofeng Wu}\email{wubf@ihep.ac.cn}\affiliation{\PKU}
\author{Chao-Guang Huang}\email{huangcg@ihep.ac.cn}\affiliation{\PKU}

%\date{\today}
\begin{abstract}
The field equations of $f(R)$ gravity  are rewritten in the form of obvious wave equations with the stress-energy
pseudotensor of the matter fields and the gravitational field as its source under the de Donder condition.
The method of multipole analysis in terms of irreducible Cartesian tensors is applied to the linearized $f(R)$ gravity, and its multipole expansion is presented explicitly.
In this expansion, the tensor part is symmetric and trace-free and is the same as that in General Relativity, and the scalar part predicts the appearance of monopole and dipole radiation in $f(R)$ gravity, as shown in literature.
As a by-product, the multipole expansion for the massive Klein-Gordon field with an external source in terms of irreducible Cartesian tensors and its corresponding stationary results are provided.
\end{abstract}
\pacs{04.50.Kd, 04.25.Nx,  04.30.-w}

\maketitle

\section{Introduction\label{Introduction}}
The recent detection of gravitational waves (GWs) by the LIGO Collaboration~\cite{TheLIGOScientific:2016agk}
is a milestone in GWs and opens new perspective in the study of General Relativity (GR) and astrophysics~\cite{Rizwana:2016qdq}.
The observations of GWs are consistent very well with GR's prediction based on the multipole expansion of gravitational
radiation, effective one-body formalism, numerical relativity, etc.

The Einstein field equations of gravity can be rewritten in the form of obvious wave
equations~\cite{Thorne:1980ru,Blanchet:2013haa} under the de Donder condition.
In this formalism, the gravitational field amplitude $h^{\mu\nu}$ is defined as
\begin{equation}\label{equ1.1}
h^{\mu\nu}:=\sqrt{-g}g^{\mu\nu}-\eta^{\mu\nu},%\tag{1.1}
\end{equation}
where $g^{\mu\nu}$ denotes the contravariant metric, $\eta^{\mu\nu}$ represents an auxiliary Minkowskian metric,
and $g$ is the determinant of metric $g_{\mu\nu}$. The source term of an obvious wave equation is the so-called stress-energy
pseudotensor of the matter fields and the gravitational field~\cite{Blanchet:2013haa}.  When $h^{\mu\nu}$ is the perturbation of
the flat metric, the field equations and the effective stress-energy tensor of GWs for the linearized GR are readily obtained.

The relativistic time-dependent massless scalar, electromagnetic, and massless tensor fields can be expanded
in terms of multipole moments~\cite{Campbell:1977jf}.  The massless scalar field is handled directly and elegantly by
using scalar spherical harmonic functions~\cite{Damour:1990gj}.  In order to deal with the vector and tensor cases
more elegantly, the method of irreducible Cartesian tensors, namely the symmetric and trace-free (STF) formalism,
has been developed~\cite{Thorne:1980ru,Blanchet:1985sp,Blanchet:1989ki}.  In Ref.~\cite{Damour:1990gj},
the STF technique is summarized, and the multipole expansions for electromagnetism and linearized GR are
presented systematically, based on STF technique. In addition to the massless fields, the multipole expansions for radiation from massive scalar and vector fields are also investigated
for periodic sources in Ref.~\cite{Krause:1994ar}.

No doubt, GR is a successful theory of gravity. Even so, it still faces many challenges to interpret many data
observed at infrared scales, which is regarded to be the signal of a breakthrough of GR at astrophysical and
cosmic scales~\cite{Capozziello:2006uv,Capozziello:2006ph,Wu:2015maa}.  An approach to deal with these
difficulties is to introduce the Extended Theories of Gravity (ETG)~\cite{Capozziello:2011et,Nojiri:2010wj},
and these theories are based on generalizations of GR.
$f(R)$ gravity~\cite{Starobinsky:1980te,Sokolowski:2007rd,Olmo:2006eh,Borowiec:2006qr} is a simple example of ETG
which modifies the Einstein-Hilbert action by adopting a general function of the Ricci scalar $R$ in the
gravitational Lagrangian.  When the ETG is introduced, the GW may possess more
polarizations~\cite{Rizwana:2016qdq}. Unfortunately, the present observation of GWs cannot make any constraint on
the non-GR polarization~\cite{TheLIGOScientific:2016sss}.
\omits{\delete
To understand the response of a detector to GW with different polarizations, it is worthwhile to explore GWs in ETG, especially in $f(R)$ gravity.}
To better understand the possible different polarizations of GWs, it is worthwhile to explore the multipole expansion for radiation
in ETC, especially in $f(R)$ gravity, in STF formalism.

For $f(R)$ gravity, besides the gravitational field amplitude $h^{\mu\nu}$ by (\ref{equ1.1}), one has to
introduce the effective gravitational field amplitude
\begin{equation}\label{equ1.2}
\tilde{h}^{\mu\nu}:=f_{R}\sqrt{-g}g^{\mu\nu}-\eta^{\mu\nu},%\tag{1.2}
\end{equation}
where $f_{R}=\partial_{R}f$. One of the purposes of the present paper is to show, by using the same method in Ref.~\cite{fockv},
that the field
equations of $f(R)$ gravity can also be rewritten in the form of obvious wave equations under the de Donder condition, and
that the source term is also the stress-energy pseudotensor of the matter fields and the gravitational field.
If $\tilde{h}^{\mu\nu}$  is a perturbation, the resulting field equations and the effective stress-energy
tensor of GWs for linearized $f(R)$ gravity are the same as the previous results given in Ref.~\cite{Berry:2011pb}.

By definitions (\ref{equ1.1}) and (\ref{equ1.2}), the true gravitational field amplitude $h^{\mu\nu}$ in $f(R)$ gravity
can be read out from $\tilde{h}^{\mu\nu}$.  For a linearized $f(R)$ gravity, $f_{R}$ depends on $aR^{(1)}$ only, where $a$ is
the coupling constant of the quadratic term in the Lagrangian of $f(R)$ gravity, and $R^{(1)}$ is the linear part of Ricci scalar $R$.
In this case, the  relation between $h^{\mu\nu}$ and $\tilde{h}^{\mu\nu}$ becomes linear
and simple, and it implies that there is a scalar part associated with the linear part of the Ricci scalar $R^{(1)}$
in the multipole expansion of linearized $f(R)$ gravity in addition to the tensor part associated
with $\tilde{h}^{\mu\nu}$.  Although the result of the tensor part is the same as that in linearized GR,
the multipole expansion of linearized $f(R)$ gravity is different from that in the linearized GR because $R^{(1)}$
satisfies massive a Klein-Gordon (KG) equation with an external source in linearized $f(R)$ gravity, which will contribute nonzero
values in the multipole expansion~\cite{Berry:2011pb,Naf:2011za,Rizwana:2016qdq,Liang:2017ahj}.
Therefore, in order to complete the multipole expansion of linearized $f(R)$ gravity, one has to
deal with the multipole expansion of $R^{(1)}$ which is described by a massive KG equation with an external source.

In Ref.~\cite{Krause:1994ar}, the multipole expansion for radiation
from massive scalar and vector fields has been obtained.  However, a general derivation of the
 multipole expansion of $R^{(1)}$ is not provided  and the main interested systems are periodic ones.
In Ref.~\cite{Naf:2011za}, the gravitational radiation in a quadratic $f(R)$ gravity has been investigated,
in which the quadratic term is treated as a small perturbation of GR, the weak-field and slow-motion approximation,
related to the Newtonian and post-Newtonian potentials, is used, and the quadratic $f(R)$ gravity is dealt with
by its analogy with scalar-tensor theories.  In both of these references, the multipole moments are not
obtained in terms of the STF technique.

In fact, since the only nontrivial term comes from $aR^{(1)}$ in a linearized $f(R)$ gravity, it is possible to show directly in $f(R)$ gravity
that the multipole expansion is valid for a large class of linearized $f(R)$ theories of gravity without any further assumption in addition to
the weak-field approximation.  The main purpose of this paper is to present the multipole expansion of $R^{(1)}$ with irreducible Cartesian tensors.
As has been shown in literature, the monopole and dipole radiation for $R^{(1)}$ do not vanish, which make GWs of $f(R)$ gravity different
from the case in GR.

The method of irreducible Cartesian tensors can also be used to study stationary cases.  Since the differential
equation satisfied by the scalar part in the linearized $f(R)$ gravity reduces to the screened Poisson equation in stationary cases,
its Green's function is the Yukawa potential.  The third purpose of this paper is
to derive the multipole expansion of the Yukawa potential for a massive scalar field in STF formalism and compare with the multipole
expansion of the Coulomb potential.

This paper is organized as follows. In Sec.~\ref{Sec:Preliminary}, we describe our notation and the
relevant formulas of STF formalism, and review the traditional formalism of $f(R)$ gravity.
In Sec.~\ref{Sec:Equation}, based on the review of how the GR is rewritten in an obvious wave
equation under de Donder condition, we show that the
field equations of $f(R)$ gravity can also be rewritten in the form of obvious wave equations with the stress-energy pseudotensor
of the matter fields and the gravitational field as their source under the de Donder condition. Upon this, we derive the
wave equation and the effective stress-energy tensor of GWs for linearized $f(R)$ gravity.
In Sec.~\ref{Sec:MultipoleExp},
we expand linearized $f(R)$ gravity by the STF multipole moments. As a by-product,
we give the multipole expansion for the massive KG field with an external source. In
Sec.~\ref{Sec:Stationary}, we discuss the stationary multipole expansion for
linearized $f(R)$ gravity and the massive KG field with an external source. In
Sec.~\ref{Sec:Conclusion}, we present the conclusions and make some discussions.
In the Appendices, we provide the detailed derivations for Eqs. (\ref{equ4.53}), (\ref{equ5.7}), and (\ref{equ5.13}), respectively.

\section{Preliminary\label{Sec:Preliminary}}

\subsection{Notation\label{Sec:Notation}}

Throughout this paper, the international system of units is used, and the signature of the metric $g_{\mu\nu}$
is $(-,+,+,+)$. The Greek indices run from 0 to 3, and the Einstein summation rule is used. When the discussion is
limited in the linearized gravitational theory and when Cartesian coordinates for flat space are used, the coordinates
$(x^{0},x^{1},x^{2},x^{3})$ are denoted by $$(ct,x_{1},x_{2},x_{3})=(ct,x_{i})$$ as though they were Minkowskian coordinates.
The Latin indices run from 1 to 3, and repeated Latin subscript indices are to be summed as though a
$\delta_{ij}$ was present. For example,
\begin{equation}\label{equ2.1}
A_{i}B_{i}=A_{1}B_{1}+A_{2}B_{2}+A_{3}B_{3}.%\tag{2.1}
\end{equation}
\noindent
The Cartesian coordinates define the spherical coordinate system $(ct,r,\theta,\varphi)$:
\begin{equation}\label{equ2.2}
x_{1}=r\sin{\theta}\cos{\varphi},\ x_{2}=r\sin{\theta}\sin{\varphi},\ x_{3}=r\cos{\theta}.%\tag{2.2}
\end{equation}

As in a flat space, the radial vector and its length are denoted by $\boldsymbol{x}$ and $r$, respectively.
The unit radial vector is $\boldsymbol{n}$, and its components are $n_{i}$, so that $n_{i}=x_{i}/r$, where $x_{i}$
are the components of $\boldsymbol{x}$. Often we shall encounter a sequence of many (say $l$) indices on a Cartesian tensor.
For ease of notation, we shall abbreviate it as follows~\cite{Thorne:1980ru}:
\begin{equation}\label{equ2.3}
B_{I_{l}}\equiv B_{i_{1}i_{2}\cdots i_{l}}.%\tag{2.3}
\end{equation}
In particular, the tensor products of $l$ radial and unit radial vectors are abbreviated by
\begin{align}
\label{equ2.5}X_{I_{l}}=X_{i_{1}i_{2}\cdots i_{l}}:= x_{i_{1}}x_{i_{2}}\cdots x_{i_{l}},\\%\tag{2.5}
\label{equ2.4}N_{I_{l}}=N_{i_{1}i_{2}\cdots i_{l}}:= n_{i_{1}}n_{i_{2}}\cdots n_{i_{l}},%\tag{2.4}\\
\end{align}
and they are related by
\begin{equation}\label{equ2.6}
  X_{I_{l}}=r^l N_{I_{l}}.
\end{equation}
The totally antisymmetric Levi-Civita tensor is denoted by $\epsilon_{ijk}$ with $\epsilon_{123}=1$.

\subsection{The relevant formulas in STF formalism \label{Sec:STFformulae}}

The relevant formulas that are useful in STF formalism are listed in the following without proof.
Given a Cartesian tensor $B_{I_{l}}$, its symmetric part is expressed by~\cite{Thorne:1980ru,Blanchet:1985sp,Damour:1990gj}
\begin{equation}\label{equ2.7}
B_{(I_{l})}=B_{(i_{1}i_{2}\cdots i_{l})}:=\frac{1}{l!}\sum_{\sigma} B_{i_{\sigma(1)}i_{\sigma(2)}\cdots i_{\sigma(l)}},%\tag{2.6}
\end{equation}
where $\sigma$ runs over all permutations of $(12\cdots l)$. The explicit STF part of $B_{I_{l}}$, denoted with a hat, is
\begin{align}
\hat{B}_{I_{l}}&\equiv B_{<I_{l}>}\equiv B_{<i_{1}i_{2}\cdots i_{l}>}\notag\\
\label{equ2.8}:&=\sum_{k=0}^{[\frac{l}{2}]}b_{k}\delta_{(i_{1}i_{2}}\cdots\delta_{i_{2k-1}i_{2k}}
S_{i_{2k+1}\cdots i_{l})a_{1}a_{1}\cdots a_{k}a_{k}},%\tag{2.7a}
\end{align}
where

\begin{align}
\label{equ2.9}b_{k}=&(-1)^{k}\frac{(2l-2k-1)!!}{(2l-1)!!}\frac{l!}{(2k)!!(l-2k)!},\\
\label{equ2.10} &S_{I_{l}}=B_{(I_{l})},%\tag{2.7c}
\end{align}
and $S_{i_{2k+1}\cdots i_l a_1a_1\cdots a_ka_k}$ represents that the latter $2k$ indices are contracted.

By the above formulas, there are~\cite{Blanchet:1985sp}
\begin{align}
\label{equ2.11}\hat{N}_{I_{l}}&=\sum_{k=0}^{[\frac{l}{2}]}b_{k}\delta_{(i_{1}i_{2}}\cdots\delta_{i_{2k-1}i_{2k}}
N_{i_{2k+1}\cdots i_{l})},\\%\tag{2.8a}
\label{equ2.12}\hat{\partial}_{I_{l}}&=\sum_{k=0}^{[\frac{l}{2}]}b_{k}\delta_{(i_{1}i_{2}}\cdots\delta_{i_{2k-1}i_{2k}}
\partial_{i_{2k+1}\cdots i_{l})},%\tag{2.8b}
\end{align}
\begin{widetext}
\begin{align}
\label{equ2.13}&\hat{\partial}_{I_{l}}f(r)=\frac{\hat{N}_{I_{l}}}{(-2)^{l}}\sum_{k=1}^{l}\frac{(-2)^{k}(2l-k-1)!}{(k-1)!(l-k)!}
r^{k-l}f^{(k)}(r),\\%\tag{2.8c}\\
\label{equ2.14}&\hat{\partial}_{I_{l}}\Big(\frac{F(t-\epsilon r/c)}{r}\Big)=(-\epsilon)^{l}\hat{N}_{I_{l}}\sum_{k=0}^{l}\frac{(l+k)!}{(2\epsilon)^{k}k!(l-k)!}
\frac{F^{(l-k)}(t-\epsilon r/c)}{c^{l-k}r^{k+1}},\qquad (\epsilon^{2}=1),%\tag{2.8d}
\end{align}
where $\partial_{I_{l}}\equiv\partial_{i_{1}i_{2}\cdots i_{l}}:=\partial_{i_{1}}\partial_{i_{2}}\cdots\partial_{i_{l}}$.

Now we will make use of the above formulas to prove
\begin{equation}\label{equ2.15}
\sum_{m'=-l}^{l}{Y^{lm'}}^*(\theta',\varphi')Y^{lm'}(\theta,\varphi)=\frac{(2l+1)!!}{4\pi l!}\hat{N}_{I_{l}}(\theta',\varphi')\hat{N}_{I_{l}}(\theta,\varphi),%\tag{2.9}
\end{equation}
\end{widetext}
where $Y^{lm'}(\theta,\varphi)$ is the spherical harmonic function, and ${Y^{lm'}}^*(\theta,\varphi)$ is its complex conjugate.
By the formula~\cite{olver2010nist}
\begin{equation}\label{equ2.16}
\sum_{m'=-l}^{l}{Y^{lm'}}^*(\theta',\varphi')Y^{lm'}(\theta,\varphi)=\frac{2l+1}{4\pi}P_{l}(\cos\tilde{\theta}),%\tag{2.10}
\end{equation}
where $P_{l}$ is the Legendre polynomial of degree $l$,
$\cos{\tilde{\theta}}=\boldsymbol{n}'\cdot\boldsymbol{n}$, and $\boldsymbol{n}',\boldsymbol{n}$ are the
unit vectors of the two directions $(\theta',\varphi')$ and $(\theta,\varphi)$, respectively.
Thus, Eqs.~(\ref{equ2.15}) and (\ref{equ2.16}) are equivalent to

\begin{equation}\label{equ2.17}
\hat{N}_{I_{l}}(\theta',\varphi')\hat{N}_{I_{l}}(\theta,\varphi)=\frac{l!}{(2l-1)!!}P_{l}(\cos\tilde{\theta}).%\tag{2.11}
\end{equation}
Since $\hat{N}_{I_{l}}(\theta,\varphi)$ is trace-free, the left-hand side of Eq. (\ref{equ2.17}) can always be written as
$$\hat{N}_{I_{l}}(\theta',\varphi')\hat{N}_{I_{l}}(\theta,\varphi)=n_{i_{1}}(\theta',\varphi')\cdots n_{i_{l}}(\theta',\varphi')\hat{N}_{I_{l}}(\theta,\varphi).$$
Further, by use of Eqs. (\ref{equ2.4}), (\ref{equ2.7}), and (\ref{equ2.11}),
\begin{widetext}
\begin{align}
\hat{N}_{I_{l}}(\theta',\varphi')\hat{N}_{I_{l}}(\theta,\varphi)&=\sum_{k=0}^{[\frac{l}{2}]}b_{k}n_{i_{1}}(\theta',\varphi')\cdots n_{i_{l}}(\theta',\varphi')
\frac{1}{l!}\sum_{\sigma} \delta_{i_{\sigma(1)}i_{\sigma(2)}}\cdots
\delta_{i_{\sigma(2k-1)}i_{\sigma(2k)}}n_{i_{\sigma(2k+1)}}(\theta,\varphi)\cdots n_{i_{\sigma(l)}}(\theta,\varphi)\notag\\
\label{equ2.18}&=\sum_{k=0}^{[\frac{l}{2}]}b_{k}(\cos\tilde{\theta})^{l-2k}.%\tag{2.12}
\end{align}
\end{widetext}
Note that
\begin{align*}
&(2k)!!=2^{k}k!,
\end{align*}

\begin{align*}
(2l-2k-1)!!=&\frac{(2l-2k)!}{(2l-2k)!!}=\frac{(2l-2k)!}{2^{l-k}(l-k)!}.
\end{align*}
Thus, from (\ref{equ2.9}), we have
\begin{align}
\label{equ2.19}b_{k}=\frac{l!}{(2l-1)!!}(-1)^{k}\frac{(2l-2k)!}{2^{l}k!(l-k)!(l-2k)!}.%\tag{2.13}
\end{align}
Inserting this result into (\ref{equ2.18}), we know that Eq.~(\ref{equ2.17}) holds by the definition of the Legendre polynomial.
\subsection{Review of $f(R)$ gravity \label{Sec:f(R)gravity}}
$f(R)$ gravity~\cite{Starobinsky:1980te,Sokolowski:2007rd,Olmo:2006eh,Borowiec:2006qr} is a generalization of GR. The action of $f(R)$ gravity is
\begin{equation}\label{equ2.20}
S=\frac{1}{2\kappa}\int dx^4\sqrt{-g}f(R)+S_{M}(g^{\mu\nu},\psi),%\tag{2.14}
\end{equation}
where $f$ is an arbitrary function of Ricci scalar $R$, $\kappa=8\pi G/c^{4}$, and $S_{M}(g^{\mu\nu},\psi)$ is the matter
action. While this action may not encode the true theory of gravity, it might contain sufficient information to
act as an effective field theory, correctly describing phenomenological behavior~\cite{Berry:2011pb,Park:2010cw}.

There are three types of $f(R)$ gravity: metric $f(R)$ gravity, Palatini $f(R)$ gravity, and metric-affine $f(R)$
gravity~\cite{Berry:2011pb}, which are not equivalent to each other \cite{Sotiriou}. We will only restrict our attention to the first one. Varying the action (\ref{equ2.20})
with respect to the metric $g^{\mu\nu}$ gives the gravitational field equations and the trace equation~\cite{Berry:2011pb,Naf:2011za}, respectively,
\begin{equation}\label{equ2.21}
H_{\mu\nu}=\kappa T_{\mu\nu},\qquad H=\kappa T,%\tag{2.15}
\end{equation}
where
\begin{align}
\label{equ2.22}H_{\mu\nu}&=-\frac{g_{\mu\nu}}{2}f+(R_{\mu\nu}+g_{\mu\nu}\square-\nabla_{\mu}\nabla_{\nu})f_{R},\\%\tag{2.16}\\
\label{equ2.23}H&=-2f+(R+3\square)f_{R},%\tag{2.17}
\end{align}
and
\begin{equation}\label{equ2.24}
T_{\mu\nu}=-\frac{2}{\sqrt{-g}}\frac{\delta S_{M}}{\delta g^{\mu\nu}}%\tag{2.18}
\end{equation}
is the stress-energy tensor of matter fields.
By Ref.~\cite{Rizwana:2016qdq}, we only consider polynomial $f(R)$ models of the form
\begin{equation}\label{equ2.25}
f(R)=R+a R^{2}+b R^{3}+\cdots,%\tag{2.19}
\end{equation}
where $a,b\cdots$ are the coupling constants, and their dimensions are $[R]^{-1},[R]^{-2}\cdots$, respectively.

%%%%%%%%%%%%%%%%%%%%%%%%%%%%%%%%%%%%%%%%%%%%%%%%%%%%%%%
%%%%%%%%%%%%%%%%%%%%%%%%%%%%%%%%%%%%%%%%%%%%%%%%%%%%%%%
%%% Wave equation & Stress-energy pseudo tensor    %%%%
%%%%%%%%%%%%%%%%%%%%%%%%%%%%%%%%%%%%%%%%%%%%%%%%%%%%%%%
%%%%%%%%%%%%%%%%%%%%%%%%%%%%%%%%%%%%%%%%%%%%%%%%%%%%%%%

\section{\uppercase{The field equations and stress-energy pseudotensor of} $f(R)$ \uppercase{gravity under de Donder condition}\label{Sec:Equation}}

\subsection{Einstein field equations under the de Donder condition \label{Sec:GRunderdeDonder}}

Firstly, we review how the Einstein field equations are
\newpage
\noindent
rewritten in the form of obvious wave equations in a fictitious flat
spacetime under the de Donder condition~\cite{Blanchet:2013haa,fockv}
\begin{align}\label{equ3.1}
\Gamma^{\alpha}&:=g^{\mu\nu}\Gamma^{\alpha}_{\mu\nu} =-\frac{1}{\sqrt{-g}}\partial_{\lambda}\overline{g}^{\lambda\alpha}=0,%\tag{3.1}
\end{align}
where
\begin{align}
\label{equ3.2}\overline{g}^{\mu\nu}&:=\sqrt{-g}g^{\mu\nu}%\tag{3.2}
\end{align}
is the densitized inverse metric. As in Refs.~\cite{Thorne:1980ru,Blanchet:2013haa}, the gravitational field amplitude $h^{\mu\nu}$ is
defined by
\begin{align}
\label{equ3.3}h^{\mu\nu}&:=\overline{g}^{\mu\nu}-\eta^{\mu\nu},%\tag{3.3}
\end{align}
where $h^{\mu\nu}$ is not necessarily a perturbation.

In Ref.~\cite{fockv}, the Ricci tensor and the Ricci scalar have been expressed by the densitized inverse metric and
the related geometrical quantities
in order to apply the de Donder condition (\ref{equ3.1}). Their expressions are
\begin{align}
\label{equ3.4}R^{\mu\nu}&=-\frac{1}{2g}\overline{g}^{\alpha\beta}\partial_{\alpha}\partial_{\beta}\overline{g}^{\mu\nu}
-\Pi^{\mu\alpha\beta}\Pi^{\nu}_{\alpha\beta}+\frac{1}{2g}\overline{g}^{\mu\nu}\overline{g}^{\alpha\beta}y_{\alpha\beta}\notag\\
&\phantom{=}+\frac{1}{2}y^{\mu}y^{\nu}+B^{\mu\nu},\\%\tag{3.4a}\\
\label{equ3.5}R&=-\frac{1}{\sqrt{-g}}\overline{g}^{\alpha\beta}y_{\alpha\beta}+B+L,%\tag{3.4b}
\end{align}
where
\begin{align*}
\label{equ3.6a}\Pi^{\mu\alpha\beta}&:=\frac{1}{2g}(\overline{g}^{\alpha\rho}\partial_{\rho}\overline{g}^{\beta\mu}
+\overline{g}^{\beta\rho}\partial_{\rho}\overline{g}^{\alpha\mu}-\overline{g}^{\mu\rho}\partial_{\rho}\overline{g}^{\alpha\beta}),\tag{3.6a}\\
\label{equ3.6b}\Pi^{\nu}_{\alpha\beta}&:=g_{\alpha\sigma}g_{\beta\rho}\Pi^{\nu\sigma\rho},\tag{3.6b}\\
\label{equ3.6c}y_{\mu}&:=\Gamma^\nu_{\mu\nu}=\partial_{\mu}\ln{\sqrt{-g}},\tag{3.6c}\\
\label{equ3.6d}y^{\mu}&:=g^{\mu\rho}y_\rho=g^{\mu\rho}\partial_{\rho}\ln{\sqrt{-g}},\tag{3.6d}\\
\label{equ3.6e}y_{\alpha\beta}&:=\partial_\alpha\Gamma^\gamma_{\beta\gamma}=\partial_{\alpha}\partial_{\beta}\ln{\sqrt{-g}},\tag{3.6e}\\
\label{equ3.6f}B^{\mu\nu}&:=\Gamma^{\mu\nu}+\frac{1}{2}(y^{\mu}\Gamma^{\nu}+y^{\nu}\Gamma^{\mu}),\tag{3.6f}\\
\label{equ3.6g}\Gamma^{\mu\nu}&:=\frac{1}{2}(g^{\mu\alpha}\partial_{\alpha}\Gamma^{\nu}+g^{\nu\alpha}
\partial_{\alpha}\Gamma^{\mu}-\Gamma^{\alpha}\partial_{\alpha}g^{\mu\nu}),\tag{3.6g}\\
\label{equ3.6h}B&:=g_{\mu\nu}B^{\mu\nu}=\Gamma+\Gamma^{\alpha}y_{\alpha},\tag{3.6h}\\
\label{equ3.6i}\Gamma&:=g_{\mu\nu}\Gamma^{\mu\nu}=\partial_{\alpha}\Gamma^{\alpha}-\frac{1}{2}g_{\mu\nu}\Gamma^{\alpha}\partial_{\alpha}
g^{\mu\nu},\tag{3.6i}\\
\label{equ3.6j}L&:=-\frac{1}{2}\Gamma^{\nu}_{\alpha\beta}\partial_{\nu}g^{\alpha\beta}-\Gamma^{\alpha}y_{\alpha}.\tag{3.6j}
\end{align*}
It should be noted that all quantities on the left-hand side of the above formulas are not tensors.
%Under the de Donder condition, $\Gamma^\alpha$, $\Gamma^{\mu\nu}$, $B^{\mu\nu}$, $\Gamma$, and $B$ vanish.
Inserting the definition of $h^{\mu\nu}$ into (\ref{equ3.4}) and (\ref{equ3.5}) gives the explicit expressions of the Ricci tensor and the Ricci scalar about the metric and $h^{\mu\nu}$, namely
\begin{widetext}
\setcounter{equation}{6}
\begin{align}
R^{\mu\nu}&=\frac{1}{2\sqrt{-g}}g^{\alpha\beta}\partial_{\alpha}\partial_{\beta}h^{\mu\nu}-\frac{1}{2g}g^{\mu\alpha}g_{\beta\tau}\partial_{\lambda}h^{\nu\tau}\partial_{\alpha}h^{\beta\lambda}+
\frac{1}{4g}g^{\mu\alpha}g^{\nu\beta}g_{\lambda\tau}g_{\epsilon\pi}\partial_{\alpha}h^{\lambda\pi}\partial_{\beta}h^{\tau\epsilon}+\frac{1}{2g}g_{\alpha\beta}g^{\lambda\tau}\partial_{\lambda}h^{\mu\alpha}\partial_{\tau}h^{\nu\beta}\notag\\
&\phantom{=}+\frac{1}{2g}\partial_{\alpha}h^{\mu\beta}\partial_{\beta}h^{\nu\alpha}-\frac{1}{2g}g^{\nu\alpha}g_{\beta\tau}\partial_{\lambda}h^{\mu\tau}\partial_{\alpha}h^{\beta\lambda}-\frac{1}{8g}g^{\mu\alpha}g^{\nu\beta}g_{\tau\epsilon}g_{\lambda\pi}\partial_{\alpha}h^{\lambda\pi}\partial_{\beta}h^{\tau\epsilon}
-\frac{1}{4g}g^{\mu\nu}g_{\rho\tau}g_{\epsilon\sigma}g^{\alpha\beta}\partial_{\alpha}h^{\rho\sigma}\partial_{\beta}h^{\tau\epsilon}\notag\\
&\phantom{=}-\frac{1}{4\sqrt{-g}}g^{\mu\nu}g^{\alpha\beta}g_{\rho\sigma}\partial_{\alpha}\partial_{\beta}h^{\rho\sigma}-\frac{1}{2\sqrt{-g}}g^{\mu\alpha}\partial_{\alpha}\partial_{\lambda}h^{\nu\lambda}-\frac{1}{2\sqrt{-g}}g^{\nu\alpha}\partial_{\alpha}\partial_{\lambda}h^{\mu\lambda}
+\frac{1}{4g}g^{\mu\nu}g_{\rho\sigma}\partial_{\alpha}h^{\rho\sigma}\partial_{\lambda}h^{\alpha\lambda}\notag\\
\label{equ3.7}&\phantom{=}-\frac{1}{2g}\partial_{\alpha}h^{\mu\nu}\partial_{\lambda}h^{\alpha\lambda},\\%\tag{3.6a}\\
%\end{align*}
%\begin{align*}
R&=-\frac{1}{4g}g^{\alpha\beta}g_{\rho\tau}g_{\epsilon\sigma}\partial_{\alpha}h^{\rho\sigma}\partial_{\beta}h^{\tau\epsilon}-\frac{1}{2\sqrt{-g}}g_{\rho\sigma}g^{\alpha\beta}\partial_{\alpha}\partial_{\beta}h^{\rho\sigma}
-\frac{1}{\sqrt{-g}}\partial_{\alpha}\partial_{\beta}h^{\alpha\beta}
+\frac{1}{2g}g_{\rho\sigma}\partial_{\alpha}h^{\rho\sigma}\partial_{\lambda}h^{\alpha\lambda}\notag\\
\label{equ3.8}&\phantom{=}-\frac{1}{2g}g_{\mu\nu}\partial_{\lambda}h^{\mu\tau}\partial_{\tau}h^{\nu\lambda}
-\frac{1}{8g}g^{\mu\nu}g_{\tau\epsilon}g_{\lambda\pi}\partial_{\mu}h^{\lambda\pi}\partial_{\nu}h^{\tau\epsilon}.%\tag{3.6b}
\end{align}
\end{widetext}
From (\ref{equ3.3}), the de Donder condition (\ref{equ3.1}) also reads
\begin{equation}\label{equ3.9}
\partial_{\mu}\overline{g}^{\mu\nu}=\partial_{\mu}h^{\mu\nu}=0.%\tag{3.7}
\end{equation}
The Ricci tensor (\ref{equ3.7}) and the Ricci scalar (\ref{equ3.8}) can be simplified by use of (\ref{equ3.9}), and, thus, the Einstein tensor
can be expressed under the de Donder condition as
\begin{align}
\label{equ3.10}G^{\mu\nu}=-\frac{1}{2g}(\square_{\eta} h^{\mu\nu}-\Lambda^{\mu\nu}_{GR}),%\tag{3.8}
\end{align}
where
$\square_{\eta}:=\eta^{\mu\nu}\partial_{\mu}\partial_{\nu}$
and
\begin{widetext}
\begin{align}
\label{equ3.11}\Lambda^{\mu\nu}_{GR}=&-h^{\alpha\beta}\partial_{\alpha}\partial_{\beta}h^{\mu\nu}
+\partial_{\alpha}h^{\mu\beta}\partial_{\beta}h^{\nu\alpha}
+\frac{1}{2}g^{\mu\nu}g_{\alpha\beta}\partial_{\lambda}h^{\alpha\tau}\partial_{\tau}h^{\beta\lambda}
-g^{\mu\alpha}g_{\beta\tau}\partial_{\lambda}h^{\nu\tau}\partial_{\alpha}h^{\beta\lambda}
-g^{\nu\alpha}g_{\beta\tau}\partial_{\lambda}h^{\mu\tau}\partial_{\alpha}h^{\beta\lambda}\notag \\
&+g_{\alpha\beta}g^{\lambda\tau}\partial_{\lambda}h^{\mu\alpha}\partial_{\tau}h^{\nu\beta}
+\frac{1}{8}(2g^{\mu\alpha}g^{\nu\beta}-g^{\mu\nu}g^{\alpha\beta})(2g_{\lambda\tau}g_{\epsilon\pi}-g_{\tau\epsilon}
g_{\lambda\pi})\partial_{\alpha}h^{\lambda\pi}\partial_{\beta}h^{\tau\epsilon}.%\tag{3.9}
\end{align}
\end{widetext}
As is clear from the above expression, $\Lambda^{\mu\nu}_{GR}$ is made of, at least, quadratic in the gravitational field amplitude $h^{\mu\nu}$ and its first and second derivatives~\cite{Blanchet:2013haa}.

Hence, the Einstein field equations
\begin{align}
\label{equ3.12} G^{\mu\nu}=\kappa T^{\mu\nu}
\end{align}
can be recast
in the form of an obvious wave equation~\cite{Thorne:1980ru,Blanchet:2013haa}
\begin{equation}\label{equ3.13}
\square_{\eta} h^{\mu\nu}=2\kappa\tau^{\mu\nu}_{GR},%\tag{3.11}
\end{equation}
where the source term
\begin{equation}\label{equ3.14}
\tau^{\mu\nu}_{GR}=|g|T^{\mu\nu}+\frac{1}{2\kappa}\Lambda^{\mu\nu}_{GR}%\tag{3.12}
\end{equation}
can be interpreted as the stress-energy pseudotensor of the matter fields and the gravitational field \cite{Blanchet:2013haa}.

If $h^{\mu\nu}$ is the perturbation, %of the metric,
the linearized field equations and the corresponding effective stress-energy
tensor of GWs are
\begin{align}
\label{equ3.15}\square_{\eta} h^{\mu\nu}&=2\kappa T^{\mu\nu},\\%\tag{3.13}\\
\label{equ3.16}t^{\mu\nu}_{GR}&=\frac{1}{2\kappa}\big<\Lambda^{\mu\nu(2)}_{GR}\big>,%\tag{3.14}
\end{align}
where $\Lambda^{\mu\nu(2)}_{GR}$ is the quadratic term of $\Lambda^{\mu\nu}_{GR}$, and $\big<\Lambda^{\mu\nu(2)}_{GR}\big>$ is its average
over a small spatial volume (several wavelengths) surrounding each point \cite{Berry:2011pb}.

%%%%%%%%%    Obvious wave equation in f(R) gravity    %%%%%%%%%

\subsection{Obvious wave equation in $f(R)$ gravity\label{Sec:f(R)underdeDonder}}

Now, we begin to rewrite the field equations of $f(R)$ gravity by using the same method.  Again,
define $h^{\mu\nu}$ as the gravitational field amplitude by (\ref{equ3.3}).
It represents the information about the metric. In order to apply the de Donder condition,
we need to define the effective gravitational field amplitude $\tilde{h}^{\mu\nu}$ by
\begin{align}
\label{equ3.17}\tilde{h}^{\mu\nu}&:=\tilde{g}^{\mu\nu}-\eta^{\mu\nu},\\%\tag{3.15a}\\
\label{equ3.18}\tilde{g}^{\mu\nu}&:=f_{R}\sqrt{-g}g^{\mu\nu}.%\tag{3.15b}
\end{align}
Obviously, besides the information of metric, it also contains the information of the function $f_{R}$,
which is, from (\ref{equ2.25}),
\begin{equation}\label{equ3.19}
f_{R}=1+2a R+3b R^{2}+\cdots.%\tag{3.16}
\end{equation}

In GR, the de Donder condition is the condition for the harmonic coordinates:
\begin{equation}\label{equ3.20}
  \square x^\mu =0.
\end{equation}
In $f(R)$ gravity, the de Donder condition in GR should be modified as
\begin{equation}\label{equ3.21}
\partial_{\mu}\tilde{g}^{\mu\nu}=\partial_{\mu}\tilde{h}^{\mu\nu}=0,%\tag{3.17}
\end{equation}
which is no longer the condition for the harmonic coordinates because it is equivalent to
\begin{equation}\label{equ3.22}
\square x^\mu =-g^{\mu\nu}\partial_\nu\ln f_R.
\end{equation}
In the following, we will prove that (\ref{equ3.21}) is indeed the generalization of the de Donder condition
in the linearized $f(R)$ gravity in Refs.~\cite{Berry:2011pb,Naf:2011za}.
For the linearized theory, $h^{\mu\nu}$ and $\tilde{h}^{\mu\nu}$ are perturbation,
and then by (\ref{equ3.2}), (\ref{equ3.3}), and (\ref{equ3.19}), there are
\begin{align}
\label{equ3.23}f_{R}&=1+2a R^{(1)}+o(h^{\mu\nu}),\\%\tag{3.18a}\\
\label{equ3.24}g&=|g_{\mu\nu}|=|\overline{g}^{\mu\nu}|=-1-h+o(h^{\mu\nu}),%\tag{3.18b}
\end{align}
where $h=\eta_{\mu\nu}h^{\mu\nu}$ is the trace of $h^{\mu\nu}$, and $o(h^{\mu\nu})$ is the higher order terms of $h^{\mu\nu}$.
Eqs.~(\ref{equ3.2}), (\ref{equ3.3}), and (\ref{equ3.24}) give
\begin{align}
\label{equ3.25}g^{\mu\nu}&=\frac{1}{\sqrt{-g}}\overline{g}^{\mu\nu}=\eta^{\mu\nu}+\overline{h}^{\mu\nu}+o(h^{\mu\nu}),\\%\tag{3.19a}\\
\label{equ3.26}g_{\mu\nu}&=\eta_{\mu\nu}-\overline{h}_{\mu\nu}+o(h^{\mu\nu}),%\tag{3.19b}
\end{align}
where
\begin{align}
\label{equ3.27}\overline{h}^{\mu\nu}&=h^{\mu\nu}-\frac{1}{2}h\eta^{\mu\nu},\\%\tag{3.20a}\\
\label{equ3.28}h&=-\overline{h}=-\eta_{\mu\nu}\overline{h}^{\mu\nu}.%\tag{3.20b}
\end{align}
It is easy to derive from (\ref{equ3.27}) and (\ref{equ3.28})
\begin{align}
\label{equ3.29}h_{\mu\nu}&=\overline{h}_{\mu\nu}-\frac{1}{2}\overline{h}\eta_{\mu\nu}.%\tag{3.21}
\end{align}
It should be emphasized that the indices of perturbation quantities $h^{\mu\nu}$ and $\overline{h}^{\mu\nu}$
are lowered by the Minkowskian metric.
(\ref{equ3.26}) means that $-\overline{h}_{\mu\nu}$ in the present paper is equivalent to the perturbation of metric $h_{\mu\nu}$ in Refs.~\cite{Berry:2011pb,Naf:2011za}.
By Eqs.~(\ref{equ3.2}), (\ref{equ3.3}), (\ref{equ3.17}), (\ref{equ3.18}), and (\ref{equ3.23}), we easily obtain
\begin{equation}\label{equ3.30}
\tilde{h}^{\mu\nu}=h^{\mu\nu}+2a R^{(1)}\eta^{\mu\nu}.%\tag{3.22}
\end{equation}
So, (\ref{equ3.21}) is equivalent to
\begin{equation}\label{equ3.31}
\partial^{\mu}h_{\mu\nu}+2a\partial_{\nu}R^{(1)}=0,%\tag{3.23}
\end{equation}
or, via (\ref{equ3.29}), to
\begin{equation}
\label{equ3.32}\partial^{\mu}\overline{h}_{\mu\nu}-\frac{1}{2}\partial_{\nu}\overline{h}+2a\partial_{\nu}R^{(1)}=0.%\tag{3.24}
\end{equation}
Eq.~(\ref{equ3.32}) is nothing but the de Donder condition
in the linearized $f(R)$ gravity, and $\overline{h}_{\mu\nu}$ in (\ref{equ3.32})
is just $-h_{\mu\nu}$ in Refs.~\cite{Berry:2011pb,Naf:2011za}.

Now we return to the (nonlinearized) $f(R)$ gravity. We will express the Ricci tensor and
the Ricci scalar in terms of the metric and $\tilde h^{\mu\nu}$ with the help of the de Donder condition (\ref{equ3.21}).
By Eqs.~(\ref{equ3.2}), (\ref{equ3.3}),
(\ref{equ3.17}), and (\ref{equ3.18}), there is
\begin{align}
\label{equ3.33}h^{\mu\nu}&=\frac{1}{f_{R}}\tilde{h}^{\mu\nu}+(\frac{1}{f_{R}}-1)\eta^{\mu\nu}.%\tag{3.25}
\end{align}
It immediately results in
\begin{align}
%\label{equ3.20b}
\label{equ3.34}\partial_{\lambda}h^{\mu\nu}&=\frac{1}{f_{R}}\partial_{\lambda}\tilde{h}^{\mu\nu}
-\overline{g}^{\mu\nu}\partial_{\lambda}\ln{f_{R}}.%\tag{3.25b}
\end{align}
With the help of the de Donder condition (\ref{equ3.21}), the substitution of (\ref{equ3.33}) and (\ref{equ3.34}) in
(\ref{equ3.7}) and (\ref{equ3.8}) gives rise to the expressions of $R^{\mu\nu}$ and $R$ in
terms of the metric and $\tilde h^{\mu\nu}$:
\begin{widetext}
\begin{align}
R^{\mu\nu}&=-\frac{1}{2f_{R}\sqrt{-g}}g^{\alpha\beta}\partial_{\alpha}\tilde{h}^{\mu\nu}\partial_{\beta}\ln{f_{R}}+
\frac{1}{2f_{R}\sqrt{-g}}g^{\alpha\beta}\partial_{\alpha}\partial_{\beta}\tilde{h}^{\mu\nu}
-\frac{1}{2}g^{\mu\nu}g^{\alpha\beta}\partial_{\alpha}\ln{f_{R}}\partial_{\beta}\ln{f_{R}}
+\frac{1}{2}g^{\mu\alpha}g^{\nu\beta}\partial_{\alpha}\ln{f_{R}}\partial_{\beta}\ln{f_{R}}\notag\\
&\phantom{=}+\frac{1}{2}g^{\mu\nu}g^{\alpha\beta}\partial_{\alpha}\partial_{\beta}\ln{f_{R}}+
g^{\mu\alpha}g^{\nu\beta}\partial_{\alpha}\partial_{\beta}\ln{f_{R}}
-\frac{1}{gf_{R}^{2}}g_{\beta\tau}g^{\alpha(\mu}\partial_{\lambda}\tilde{h}^{\nu)\tau}\partial_{\alpha}\tilde{h}^{\beta\lambda}
+\frac{1}{4gf_{R}^{2}}g_{\lambda\tau}g_{\epsilon\pi}g^{\mu\alpha}g^{\nu\beta}\partial_{\alpha}\tilde{h}^{\lambda\pi}
\partial_{\beta}\tilde{h}^{\tau\epsilon}\notag\\
&\phantom{=}-\frac{1}{2f_{R}\sqrt{-g}}g_{\rho\sigma}g^{\mu(\alpha}g^{\beta)\nu}\partial_{\alpha}\tilde{h}^{\rho\sigma}\partial_{\beta}\ln{f_{R}}
+\frac{1}{2gf_{R}^{2}}g_{\alpha\beta}g^{\lambda\tau}\partial_{\lambda}\tilde{h}^{\mu\alpha}\partial_{\tau}\tilde{h}^{\nu\beta}
+\frac{1}{2gf_{R}^{2}}\partial_{\alpha}\tilde{h}^{\mu\beta}\partial_{\beta}\tilde{h}^{\nu\alpha}\notag\\
&\phantom{=}+\frac{1}{f_{R}\sqrt{-g}}g^{\alpha(\mu}\partial_{\alpha}\tilde{h}^{\nu)\beta}\partial_{\beta}\ln{f_{R}}
-\frac{1}{8gf_{R}^{2}}g_{\tau\epsilon}g_{\lambda\pi}g^{\mu\alpha}g^{\nu\beta}\partial_{\alpha}\tilde{h}^{\lambda\pi}
\partial_{\beta}\tilde{h}^{\tau\epsilon}
-\frac{1}{4gf_{R}^{2}}g_{\rho\tau}g_{\epsilon\sigma}g^{\mu\nu}g^{\alpha\beta}\partial_{\alpha}\tilde{h}^{\rho\sigma}
\partial_{\beta}\tilde{h}^{\tau\epsilon}\notag\\
\label{equ3.35}&\phantom{=}-\frac{1}{4f_{R}\sqrt{-g}}g^{\mu\nu}g^{\alpha\beta}g_{\rho\sigma}\partial_{\alpha}\partial_{\beta}\tilde{h}^{\rho\sigma}
+\frac{1}{4f_{R}\sqrt{-g}}g^{\mu\nu}g^{\alpha\beta}g_{\rho\sigma}\partial_{\alpha}\tilde{h}^{\rho\sigma}\partial_{\beta}\ln{f_{R}},%\tag{3.27a}\\
\end{align}
\begin{align}
R&=-\frac{1}{4gf_{R}^{2}}g_{\rho\tau}g_{\epsilon\sigma}g^{\alpha\beta}\partial_{\alpha}\tilde{h}^{\rho\sigma}\partial_{\beta}\tilde{h}^{\tau\epsilon}
-\frac{3}{2}g^{\alpha\beta}\partial_{\alpha}\ln{f_{R}}\partial_{\beta}\ln{f_{R}}
-\frac{1}{2f_{R}\sqrt{-g}}g_{\rho\sigma}g^{\alpha\beta}\partial_{\alpha}\partial_{\beta}\tilde{h}^{\rho\sigma}
+3g^{\alpha\beta}\partial_{\alpha}\partial_{\beta}\ln{f_{R}}\notag\\
\label{equ3.36}&\phantom{=}-\frac{1}{2gf_{R}^{2}}g_{\alpha\beta}\partial_{\lambda}\tilde{h}^{\alpha\tau}\partial_{\tau}\tilde{h}^{\beta\lambda}
-\frac{1}{8gf_{R}^{2}}g^{\alpha\beta}g_{\tau\epsilon}g_{\lambda\pi}\partial_{\alpha}\tilde{h}^{\lambda\pi}\partial_{\beta}\tilde{h}^{\tau\epsilon}.%\tag{3.27b}
\end{align}
\end{widetext}
We begin to consider the field equations (\ref{equ2.21}) for $f(R)$ gravity. Eq.~(\ref{equ2.22}) can be split
into two parts,
\begin{align}
\label{equ3.37}H^{\mu\nu}&:=H^{\mu\nu}_{1}+H^{\mu\nu}_{2},%\tag{3.28a}\\
\end{align}
where
\begin{align}
\label{equ3.38}H^{\mu\nu}_{1}&:=-\frac{1}{2}g^{\mu\nu}f+R^{\mu\nu}f_{R},\\%\tag{3.28b}\\
\label{equ3.39}H^{\mu\nu}_{2}&:=(g^{\mu\nu}\square-\nabla^{\mu}\nabla^{\nu})f_{R}.%\tag{3.28c}
\end{align}
For Lagrangian (\ref{equ2.25}), Eq.~(\ref{equ3.38}) reads \medskip
\\
\begin{align}
\label{equ3.40}
H^{\mu\nu}_{1}=&G^{\mu\nu}-\frac{a}{2}g^{\mu\nu}R^{2}+2aR^{\mu\nu}R-\frac{b}{2}g^{\mu\nu}R^{3}+3bR^{\mu\nu}R^{2} \notag\\
&+\mbox{higher order terms}.
\end{align}
By (\ref{equ3.35}) and (\ref{equ3.36}),  the expression of Einstein tensor $G^{\mu\nu}$ in terms of the metric and $\tilde{h}^{\mu\nu}$ can be obtained,
\begin{widetext}
\begin{align}
G^{\mu\nu}&=-\frac{1}{2f_{R}\sqrt{-g}}g^{\alpha\beta}\partial_{\alpha}\tilde{h}^{\mu\nu}\partial_{\beta}\ln{f_{R}}+
\frac{1}{2f_{R}\sqrt{-g}}g^{\alpha\beta}\partial_{\alpha}\partial_{\beta}\tilde{h}^{\mu\nu}
+\frac{1}{4}g^{\mu\nu}g^{\alpha\beta}\partial_{\alpha}\ln{f_{R}}\partial_{\beta}\ln{f_{R}}
+\frac{1}{2}g^{\mu\alpha}g^{\nu\beta}\partial_{\alpha}\ln{f_{R}}\partial_{\beta}\ln{f_{R}}\notag\\
&\phantom{=}-g^{\mu\nu}g^{\alpha\beta}\partial_{\alpha}\partial_{\beta}\ln{f_{R}}+
g^{\mu\alpha}g^{\nu\beta}\partial_{\alpha}\partial_{\beta}\ln{f_{R}}
-\frac{1}{gf_{R}^{2}}g_{\beta\tau}g^{\alpha(\mu}\partial_{\lambda}\tilde{h}^{\nu)\tau}\partial_{\alpha}\tilde{h}^{\beta\lambda}
-\frac{1}{2f_{R}\sqrt{-g}}g_{\rho\sigma}g^{\mu(\alpha}g^{\beta)\nu}\partial_{\alpha}\tilde{h}^{\rho\sigma}\partial_{\beta}\ln{f_{R}}\notag\\
&\phantom{=}+\frac{1}{2gf_{R}^{2}}g_{\alpha\beta}g^{\lambda\tau}\partial_{\lambda}\tilde{h}^{\mu\alpha}\partial_{\tau}\tilde{h}^{\nu\beta}
+\frac{1}{2gf_{R}^{2}}\partial_{\alpha}\tilde{h}^{\mu\beta}\partial_{\beta}\tilde{h}^{\nu\alpha}
+\frac{1}{f_{R}\sqrt{-g}}g^{\alpha(\mu}\partial_{\alpha}\tilde{h}^{\nu)\beta}\partial_{\beta}\ln{f_{R}}
+\frac{1}{4gf_{R}^{2}}g^{\mu\nu}g_{\alpha\beta}\partial_{\lambda}\tilde{h}^{\alpha\tau}
\partial_{\tau}\tilde{h}^{\beta\lambda} \notag \\
\label{equ3.41}&\phantom{=}+\frac{1}{4f_{R}\sqrt{-g}}g^{\mu\nu}g^{\alpha\beta}g_{\rho\sigma}
\partial_{\alpha}\tilde{h}^{\rho\sigma}\partial_{\beta}\ln{f_{R}}
+\frac{1}{16gf_{R}^{2}}(2g^{\mu\alpha}g^{\nu\beta}-g^{\mu\nu}g^{\alpha\beta})(2g_{\lambda\tau}g_{\epsilon\pi}
-g_{\epsilon\tau}g_{\lambda\pi})\partial_{\alpha}\tilde{h}^{\lambda\pi}\partial_{\beta}\tilde{h}^{\tau\epsilon}.%\tag{3.30}
\end{align}
\end{widetext}
When $f(R)$ gravity reduces to GR, namely,
\begin{equation}
\label{equ3.42}f(R)=R, %\tag{3.31}
\end{equation}
then
\begin{align}
\label{equ3.43}f_{R}&=1,\\%\tag{3.32a}\\
\label{equ3.44}\tilde{h}^{\mu\nu}&=h^{\mu\nu}%\tag{3.32b}
\end{align}
by (\ref{equ3.19}) and (\ref{equ3.33}), and Eq.~(\ref{equ3.41}) reduces to the expression
of Einstein tensor, namely (\ref{equ3.10}) in GR.

For scalar $f_{R}$,
%\begin{align*}
\[
\nabla^{\mu}\nabla^{\nu}f_{R}=g^{\mu\alpha}g^{\nu\beta}\partial_{\alpha}\partial_{\beta}f_{R}-\Gamma^{\lambda\mu\nu}\partial_{\lambda}f_{R},\\
\]
\[
\square f_{R}=g^{\alpha\beta}\partial_{\alpha}\partial_{\beta}f_{R}-\Gamma^{\lambda}\partial_{\lambda}f_{R},
\]
%\end{align*}
and then (\ref{equ3.39}) reads
\begin{align}
H^{\mu\nu}_{2}&=(g^{\mu\nu}g^{\alpha\beta}-g^{\mu\alpha}g^{\nu\beta})\partial_{\alpha}\partial_{\beta}f_{R}\notag\\
\label{equ3.45}&\phantom{=}-g^{\mu\nu}\Gamma^{\lambda}\partial_{\lambda}f_{R}+\Gamma^{\lambda\mu\nu}\partial_{\lambda}f_{R},%\tag{3.33}
\end{align}
where
\begin{align}
\label{equ3.46}\Gamma^{\lambda\mu\nu}:&=g^{\mu\alpha}g^{\nu\beta}\Gamma^{\lambda}_{\alpha\beta}\notag\\
&=\Pi^{\lambda\mu\nu}+\frac{1}{2}(y^{\mu}g^{\nu\lambda}+y^{\nu}g^{\mu\lambda}-y^{\lambda}g^{\mu\nu})%\tag{3.34}
\end{align}
by Ref.~\cite{fockv}.  By use of (\ref{equ3.2}), (\ref{equ3.3}), (\ref{equ3.6a}), (\ref{equ3.6d}), (\ref{equ3.21}), (\ref{equ3.33}),
and (\ref{equ3.34}), Eq.~(\ref{equ3.46}) can be expressed as
\begin{widetext}
\begin{align}
\Gamma^{\lambda\mu\nu}=&-\frac{1}{f_{R}\sqrt{-g}}g^{\rho(\mu}\partial_{\rho}\tilde{h}^{\nu)\lambda}+
\frac{1}{2f_{R}\sqrt{-g}}g^{\lambda\rho}\partial_{\rho}\tilde{h}^{\mu\nu}
+\frac{1}{2f_{R}\sqrt{-g}}g_{\alpha\beta}g^{\rho(\mu}g^{\nu)\lambda}\partial_{\rho}\tilde{h}^{\alpha\beta}
-\frac{1}{4f_{R}\sqrt{-g}}g_{\alpha\beta}g^{\mu\nu}g^{\lambda\rho}\partial_{\rho}\tilde{h}^{\alpha\beta} \notag\\
\label{equ3.47}\phantom{=}&-g^{\rho(\mu}g^{\nu)\lambda}\partial_{\rho}\ln{f_{R}}
+\frac{1}{2}g^{\mu\nu}g^{\lambda\rho}\partial_{\rho}\ln{f_{R}}.%\tag{3.35}
\end{align}
Therefore,
\begin{align}
H^{\mu\nu}_{2}=&f_{R}g^{\mu\nu}g^{\alpha\beta}\partial_{\alpha}\partial_{\beta}\ln{f_{R}}-
f_{R}g^{\mu\alpha}g^{\nu\beta}\partial_{\alpha}\partial_{\beta}\ln{f_{R}}-f_{R}g^{\mu\alpha}g^{\nu\beta}\partial_{\alpha}\ln{f_{R}}\partial_{\beta}\ln{f_{R}}
-\frac{1}{\sqrt{-g}}g^{\rho(\mu}\partial_{\rho}\tilde{h}^{\nu)\lambda}\partial_{\lambda}\ln{f_{R}}\notag\\
\phantom{=}&+
\frac{1}{2\sqrt{-g}}g^{\lambda\rho}\partial_{\rho}\tilde{h}^{\mu\nu}\partial_{\lambda}\ln{f_{R}}
+\frac{1}{2\sqrt{-g}}g_{\alpha\beta}g^{\rho(\mu}g^{\nu)\lambda}\partial_{\rho}\tilde{h}^{\alpha\beta}\partial_{\lambda}\ln{f_{R}}
-\frac{1}{4\sqrt{-g}}g_{\alpha\beta}g^{\mu\nu}g^{\lambda\rho}\partial_{\rho}\tilde{h}^{\alpha\beta}\partial_{\lambda}\ln{f_{R}}\notag\\
\label{equ3.48}\phantom{=}&-f_{R}g^{\rho(\mu}g^{\nu)\lambda}\partial_{\rho}\ln{f_{R}}\partial_{\lambda}\ln{f_{R}}
+\frac{1}{2}f_{R}g^{\mu\nu}g^{\lambda\rho}\partial_{\rho}\ln{f_{R}}\partial_{\lambda}\ln{f_{R}}.%\tag{3.36}
\end{align}
The combination of (\ref{equ3.37}), (\ref{equ3.40}), and (\ref{equ3.48}) brings about the expression of $H^{\mu\nu}$:
\begin{align}
\label{equ3.49}H^{\mu\nu}=-\frac{1}{2gf_{R}^{2}}(\square_{\eta} \tilde{h}^{\mu\nu}-\Lambda^{\mu\nu}_{f}),%\tag{3.37}
\end{align}
where
\begin{align}
\label{equ3.50}\Lambda^{\mu\nu}_{f}=&-\tilde{h}^{\alpha\beta}\partial_{\alpha}\partial_{\beta}\tilde{h}^{\mu\nu}
-(f_{R}-1)\tilde{g}^{\alpha\beta}\partial_{\alpha}\tilde{h}^{\mu\nu}\partial_{\beta}\ln{f_{R}}
-\frac{1}{2}(1+2f_{R})\tilde{g}^{\mu\nu}\tilde{g}^{\alpha\beta}\partial_{\alpha}\ln{f_{R}}\partial_{\beta}\ln{f_{R}}\notag\\
&-(1-4f_{R})\tilde{g}^{\mu\alpha}\tilde{g}^{\nu\beta}\partial_{\alpha}\ln{f_{R}}\partial_{\beta}\ln{f_{R}}
-2(f_{R}-1)\tilde{g}^{\mu\nu}\tilde{g}^{\alpha\beta}\partial_{\alpha}\partial_{\beta}\ln{f_{R}}
+2(f_{R}-1)\tilde{g}^{\mu\alpha}\tilde{g}^{\nu\beta}\partial_{\alpha}\partial_{\beta}\ln{f_{R}}\notag\\
&-2\tilde{g}_{\beta\tau}\tilde{g}^{\alpha(\mu}\partial_{\lambda}\tilde{h}^{\nu)\tau}\partial_{\alpha}\tilde{h}^{\beta\lambda}
-(f_{R}-1)\tilde{g}_{\rho\sigma}\tilde{g}^{\mu(\alpha}\tilde{g}^{\beta)\nu}\partial_{\alpha}\tilde{h}^{\rho\sigma}\partial_{\beta}\ln{f_{R}}
+\tilde{g}_{\alpha\beta}\tilde{g}^{\lambda\tau}\partial_{\lambda}\tilde{h}^{\mu\alpha}\partial_{\tau}\tilde{h}^{\nu\beta}\notag\\
&+\partial_{\alpha}\tilde{h}^{\mu\beta}\partial_{\beta}\tilde{h}^{\nu\alpha}
-2(1-f_{R})\tilde{g}^{\alpha(\mu}\partial_{\alpha}\tilde{h}^{\nu)\beta}\partial_{\beta}\ln{f_{R}}
-\frac{1}{2}(1-f_{R})\tilde{g}_{\rho\sigma}\tilde{g}^{\mu\nu}\tilde{g}^{\alpha\beta}\partial_{\alpha}\tilde{h}^{\rho\sigma}\partial_{\beta}\ln{f_{R}}\notag\\
&+\frac{1}{2}\tilde{g}_{\alpha\beta}\tilde{g}^{\mu\nu}\partial_{\lambda}\tilde{h}^{\alpha\tau}\partial_{\tau}\tilde{h}^{\beta\lambda}
+\frac{1}{8}(2\tilde{g}^{\mu\alpha}\tilde{g}^{\nu\beta}
-\tilde{g}^{\mu\nu}\tilde{g}^{\alpha\beta})(2\tilde{g}_{\lambda\tau}\tilde{g}_{\epsilon\pi}-\tilde{g}_{\epsilon\tau}\tilde{g}_{\lambda\pi})
\partial_{\alpha}\tilde{h}^{\lambda\pi}\partial_{\beta}\tilde{h}^{\tau\epsilon}\notag\\
&+a\sqrt{-g}f_{R}\tilde{g}^{\mu\nu}R^{2}+b\sqrt{-g}f_{R}\tilde{g}^{\mu\nu}R^{3}+4agf_{R}^{2}R^{\mu\nu}R+6bgf_{R}^{2}R^{\mu\nu}R^{2}+\mbox{higher order terms},%\tag{3.38}
\end{align}
\end{widetext}
and
\begin{align}
\label{equ3.51}\tilde{g}_{\mu\nu}:=\frac{1}{\sqrt{-g}f_{R}}g_{\mu\nu}%\tag{3.39}
\end{align}
is the inverse of $\tilde{g}^{\mu\nu}$, namely, there is
\begin{align}
\label{equ3.52}\tilde{g}_{\mu\lambda}\tilde{g}^{\lambda\nu}=\delta^{\nu}_{\mu}.%\tag{3.40}
\end{align}
It is easy to check that $\Lambda^{\mu\nu}_{f}$, just like $\Lambda^{\mu\nu}_{GR}$, is also made of, at least, quadratic in the effective gravitational
field amplitude $\tilde{h}^{\mu\nu}$ and its first and second derivatives.
Moreover, $\Lambda^{\mu\nu}_{f}$ can reduce to the expression of $\Lambda^{\mu\nu}_{GR}$ in (\ref{equ3.11}),
 when $f(R)$ gravity reduces to GR by (\ref{equ3.43}) and (\ref{equ3.44}). According to Eq.~(\ref{equ3.49}), the field equations of $f(R)$ gravity are transformed into the form of an obvious wave equation
\begin{equation}\label{equ3.53}
\square_{\eta} \tilde{h}^{\mu\nu}=2\kappa\tau^{\mu\nu}_{f}%\tag{3.41}
\end{equation}
under the de Donder condition, where
the source term
\begin{equation}\label{equ3.54}
\tau^{\mu\nu}_{f}=|g|f_{R}^{2}T^{\mu\nu}+\frac{1}{2\kappa}\Lambda^{\mu\nu}_{f}%\tag{3.42}
\end{equation}
is the stress-energy pseudotensor of the matter fields and the gravitational field.

%%%%%%%%  Linearized f(R) gravity *************

\subsection{Linearized $f(R)$ gravity}
If $\tilde{h}^{\mu\nu}$ is a perturbation, namely,
\begin{equation}\label{equ3.55}
|\tilde{h}^{\mu\nu}|\ll1,%\tag{3.43}
\end{equation}
the linearized gravitational field equations is
\begin{align}
\label{equ3.56}\square_{\eta}\tilde{h}&^{\mu\nu}=2\kappa T^{\mu\nu}%\tag{3.44}
\end{align}
by (\ref{equ3.53}) and (\ref{equ3.54}). Eq.~(\ref{equ3.56}) is the basis of multipole expansion with irreducible
Cartesian tensors.

In order to derive the effective stress-energy tensor of GWs for the linearized $f(R)$ gravity, we still need some formulas.
Firstly by (\ref{equ3.17}) and (\ref{equ3.52}), we have
\begin{equation}\label{equ3.57}
\tilde{g}_{\mu\nu}=\eta_{\mu\nu}-\tilde{h}_{\mu\nu}.%\tag{3.45}
\end{equation}
Next, by (\ref{equ3.23}), (\ref{equ3.24}), and (\ref{equ3.30}),  there are
\begin{align}
\label{equ3.58}\sqrt{-g}f_{R}&=1+\frac{\tilde{h}}{2}-2aR^{(1)}+o(\tilde{h}^{\mu\nu}),\\%\tag{3.47a}\\
\label{equ3.59}-gf_{R}^{2}&=1+\tilde{h}-4aR^{(1)}+o(\tilde{h}^{\mu\nu}).%\tag{3.47b}
\end{align}
The substitution of (\ref{equ3.17}), (\ref{equ3.23}), (\ref{equ3.57}), (\ref{equ3.58}), and (\ref{equ3.59}) in (\ref{equ3.50}) gives rise to
%\begin{widetext}%\setcounter{equation}{47}
\begin{align}\label{equ3.60}
\Lambda^{\mu\nu(2)}_{f}&=\Lambda^{\mu\nu(2)}_{GR}(\tilde h^{\alpha\beta})+a\eta^{\mu\nu}{R^{(1)}}^2-4aR^{\mu\nu(1)}R^{(1)}\notag\\
&-6a^{2}\eta^{\mu\nu}\partial^{\alpha}R^{(1)}\partial_{\alpha}R^{(1)}
+12a^{2}\partial^{\mu}R^{(1)}\partial^{\nu}R^{(1)}\notag\\
&-8a^{2}\eta^{\mu\nu}R^{(1)}\square_{\eta} R^{(1)}
+8a^{2}R^{(1)}\partial^{\mu}\partial^{\nu}R^{(1)},%\tag{3.48}
\end{align}
%\end{widetext}
where $\Lambda^{\mu\nu(2)}_{f}$ is the quadratic term of $\Lambda^{\mu\nu}_{f}$, and $\Lambda^{\mu\nu(2)}_{GR}(\tilde h^{\alpha\beta})$ is the
quadratic term of  $\Lambda^{\mu\nu}_{GR}$ with the replace of the variables $h^{\alpha\beta}$ by $\tilde h^{\alpha\beta}$.

We need to simplify $\Lambda^{\mu\nu(2)}_{f}$. By use of (\ref{equ3.23})---(\ref{equ3.26}), Eq.~(\ref{equ3.36})
 reduces to
\begin{align}
\label{equ3.61}R^{(1)}=6a\square_{\eta} R^{(1)}-\frac{1}{2}\square_{\eta}\tilde{h}. %\tag{3.49}
\end{align}
Furthermore, (\ref{equ2.21}) and (\ref{equ3.49}) lead to
\begin{align}
\label{equ3.62}H^{(1)}=\frac{1}{2}\square_{\eta}\tilde{h}=\kappa T^{(1)}, %\tag{3.50}
\end{align}
where $T^{(1)}$ is the linearized term of the trace of $T^{\mu\nu}$.  Eqs.~(\ref{equ3.61}) and (\ref{equ3.62}) imply
\begin{align}
\label{equ3.63}\square_{\eta} R^{(1)}-m^{2}R^{(1)}=m^{2}\kappa T^{(1)},%\tag{3.51a}
\end{align}
where
\begin{align}
\label{equ3.64}m^{2}:=\frac{1}{6a},%\tag{3.51b}
\end{align}
which shows that $R^{(1)}$ satisfies a massive KG equation with an external source, as shown in the literature~\cite{Naf:2011za}.
Therefore, by Eqs.~(\ref{equ3.21}) and (\ref{equ3.56}), the complete linearized equations of $f(R)$ gravity outside sources are
\begin{align}
\label{equ3.65}\square_{\eta} R^{(1)}&=\frac{1}{6a}R^{(1)},\\%\tag{3.52a}\\
\label{equ3.66}\square_{\eta} \tilde{h}^{\mu\nu}&=0,\\%\tag{3.52b}\\
\label{equ3.67}\partial_{\mu}\tilde{h}^{\mu\nu}&=0.%\tag{3.52c}
\end{align}

Now, we will take the average $\big<\cdots\big>$ over a small spatial volume (several wavelengths)
surrounding each point again.
The relevant rules for the average are~\cite{Berry:2011pb}
\begin{align}
\label{equ3.68}\big<\partial_{\mu}X\big>&=0,\\%\tag{3.53a}\\
\label{equ3.69}\big<A(\partial_{\mu}B)\big>&=-\big<(\partial_{\mu}A)B\big>,%\tag{3.53b}
\end{align}
where $X,A,B$ are three arbitrary quantities.  By use of Eqs.~(\ref{equ3.65})---(\ref{equ3.69}), the average of
(\ref{equ3.60}) outside the sources reduces to
\begin{align}
\label{equ3.70}\big<\Lambda^{\mu\nu(2)}_{f}\big>=\big<\Lambda^{\mu\nu(2)}_{GR}(\tilde h^{\alpha\beta})\big>+12a^{2}\big<\partial^{\mu}R^{(1)}\partial^{\nu}R^{(1)}\big>.%\tag{3.54}
\end{align}
It gives the effective stress-energy tensor of GWs in linearized $f(R)$ gravity,
\begin{align}
\label{equ3.71}t^{\mu\nu}_{f}:=\frac{1}{2\kappa}\big<\Lambda^{\mu\nu(2)}_{f}\big>
=t^{\mu\nu}_{GR}(\tilde h^{\alpha\beta})+\frac{6a^{2}}{\kappa}\big<\partial^{\mu}R^{(1)}\partial^{\nu}R^{(1)}\big>.%\tag{3.55}
\end{align}
%\end{widetext}
It should be noted that the result does not depend on $b$ and higher order coupling constants. It is easy to prove that the linearized field equations
(\ref{equ3.56}) and the effective stress-energy tensor of GWs (\ref{equ3.71}) are the same as the previous results given
in Ref.~\cite{Berry:2011pb}.

%%%%%%%%%%%%%%%%%%%%%%%%%%%%%%%%%%%%%%%%%%%%%%%%%%%%%%%
%%%%%%%%%%%%%%%%%%%%%%%%%%%%%%%%%%%%%%%%%%%%%%%%%%%%%%%
%%%           Multipole Expansion                  %%%%
%%%%%%%%%%%%%%%%%%%%%%%%%%%%%%%%%%%%%%%%%%%%%%%%%%%%%%%
%%%%%%%%%%%%%%%%%%%%%%%%%%%%%%%%%%%%%%%%%%%%%%%%%%%%%%%

\section{\uppercase{The multipole expansion of linearized} $f(R)$ \uppercase{gravity}\label{Sec:MultipoleExp}}
In this section, we will discuss the multipole expansion of linearized $f(R)$ gravity with the irreducible Cartesian tensors.
According to the preceding section, $\tilde{h}^{\mu\nu}$ is only the effective gravitational field amplitude.
It contains the information of function $f_{R}$ in addition to the information of metric.
On the other hand, the gravitational amplitude $h^{\mu\nu}$ carries all pure information of metric. It implies that
we need to find out the multipole
expansion of $h^{\mu\nu}$.  Since the relationship between $h^{\mu\nu}$ and $\tilde{h}^{\mu\nu}$ is given by (\ref{equ3.30}) or
\begin{equation}
\label{equ4.1}h^{\mu\nu}=\tilde{h}^{\mu\nu}-2aR^{(1)}\eta^{\mu\nu},%\tag{4.1}
\end{equation}
the multipole expansions of tensor part $\tilde{h}^{\mu\nu}$ and of the scalar part associated with $R^{(1)}$ should be dealt with separately.
As is well known, it is the scalar part that makes the
multipole expansion of linearized $f(R)$ gravity different from that of GR.\medskip%

%%%%%%%%%%%%%   expansion for h^{\mu\nu}  %%%%%%%%%%%%%%%

\subsection{The multipole expansion of $\tilde{h}^{\mu\nu}$\label{Sec:MEtensorpart}}
The linearized field equations (\ref{equ3.15}) and (\ref{equ3.56}) and the de Donder conditions (\ref{equ3.9}) and (\ref{equ3.21}) show that $h^{\mu\nu}$
in linearized GR and $\tilde{h}^{\mu\nu}$ in linearized $f(R)$ gravity satisfy the same wave equation and the same gauge
condition. So, according to Refs.~\cite{Blanchet:1985sp,Damour:1990gj}, the multipole expansion of $\tilde{h}^{\mu\nu}$
in linearized $f(R)$ are the same as that of $h^{\mu\nu}$ in linearized GR, namely,
\begin{widetext}
\begin{equation}\label{equ4.2}
\left\{\begin{array}{l}
\displaystyle\tilde{h}^{00}(t,\boldsymbol{x})= -\frac{4G}{c^{2}}\sum_{l=0}^{\infty}\frac{(-1)^{l}}{l!}\partial_{I_{l}}\left( \frac{\hat{M}_{I_{l}}(u)}{r}\right),\smallskip\\
\displaystyle\tilde{h}^{0i}(t,\boldsymbol{x})= \frac{4G}{c^{3}}\sum_{l=1}^{\infty}\frac{(-1)^{l}}{l!}\partial_{I_{l-1}}\left(\frac{\partial_{t}\hat{M}_{iI_{l-1}}(u)}{r}\right) +\frac{4G}{c^{3}}\sum_{l=1}^{\infty}\frac{(-1)^{l}l}{(l+1)!}\epsilon_{iab}\partial_{aI_{l-1}}\left( \frac{\hat{S}_{bI_{l-1}}(u)}{r}\right),\smallskip\\
\displaystyle\tilde{h}^{ij}(t,\boldsymbol{x})=-\frac{4G}{c^{4}}\sum_{l=2}^{\infty}\frac{(-1)^{l}}{l!}\partial_{I_{l-2}}\left(\frac{\partial_{t}^{2}\hat{M}_{ijI_{l-2}}(u)}{r}\right) -\frac{8G}{c^{4}}\sum_{l=2}^{\infty}\frac{(-1)^{l}l}{(l+1)!}\partial_{aI_{l-2}}\left(\frac{\epsilon_{ab(i}\partial_{t}\hat{S}_{j)bI_{l-2}}(u)}{r}\right),
\end{array}\right.%\tag{4.2}
\end{equation}
where
\begin{equation}\label{equ4.3}
\left\{\begin{array}{l}
\displaystyle\hat{M}_{I_{l}}(u)=\frac{1}{c^{2}}\int d^{3}x'\left( \hat{X'}_{I_{l}}\left(\overline{T}^{00}_{l}(u,\boldsymbol{x}')+\overline{T}^{aa}_{l}(u,\boldsymbol{x}')\right)
-\frac{4(2l+1)}{c(l+1)(2l+3)}\hat{X'}_{aI_{l}}\partial_{t}\overline{T}^{0a}_{l+1}(u,\boldsymbol{x}') \right . \\
\displaystyle\qquad\qquad\qquad\qquad\qquad\qquad\qquad\left .
+\frac{2(2l+1)}{c^2(l+1)(l+2)(2l+5)}\hat{X'}_{abI_{l}}\partial_{t}^{2}\overline{T}^{ab}_{l+2}(u,\boldsymbol{x}')\right ),\smallskip\\
\displaystyle\hat{S}_{I_{l}}(u)=\frac{1}{c}\int d^{3}x'\left(\epsilon_{ab<i_{1}}\hat{X'}_{|a|i_{2}\cdots i_{l}>}\overline{T}^{0b}_{l}(u,\boldsymbol{x}') \right . \\
\displaystyle\qquad\qquad\qquad\qquad\qquad\qquad\qquad\left . -\frac{2l+1}{c(l+2)(2l+3)}\epsilon_{ab<i_{1}}\hat{X'}_{|ac|i_{2}\cdots i_{l}>}\partial_{t}\overline{T}^{cb}_{l+1}(u,\boldsymbol{x}')\right),\ l\geq1
\end{array}\right.%\tag{4.3}
\end{equation}
are referred to as the mass-type and current-type source multipole moments, respectively~\cite{Blanchet:2013haa},
$u=t-r/c$ is the retarded time, $\partial_t^2$ is the second derivative with respect to $t$, the symbol $<i_{1}|a|i_{2}\cdots i_{l}>$ and $<i_{1}|ac|i_{2}\cdots i_{l}>$
represent that $a$ and $c$ are not STF indices, and \cite{Damour:1990gj}
\begin{align}
\label{equ4.4}\overline{T}^{\mu\nu}_{l}(u,\boldsymbol{x}'):=\frac{(2l+1)!!}{2^{l+1}l!}\int_{-1}^{1}(1-z^2)^{l}T^{\mu\nu}\Big(u+\frac{zr'}{c},
\boldsymbol{x}'\Big)dz.%\tag{4.4}
\end{align}
%\end{widetext}
%Although we have obtained the multipole expansion of $\tilde{h}^{\mu\nu}$, (\ref{equ3.19}) implies that if we
%want to solve the complete multipole expansion of linearized $f(R)$ gravity, how to solve the multipole expansion of $R^{(1)}$ is the
%key difficulty we face.

%%%%%%%%%%%%%   expansion for R^{(1)}  %%%%%%%%%%%%%%%

\subsection{The multipole expansion of $R^{(1)}$\label{Sec:MEscalarpart}}
Eq.~(\ref{equ3.63}) shows that $R^{(1)}$ satisfies the massive KG equation with an external source.
The following key task is to
find out the multipole expansion for the massive KG field with irreducible Cartesian tensors.
We will follow the method in Ref.~\cite{Campbell:1977jf}
to solve this problem.
The retarded Green's function of (\ref{equ3.63}) is~\cite{Naf:2011za}
%\begin{widetext}
\begin{align}
\label{equ4.5}\mathcal{G}(t,\boldsymbol{x};t',\boldsymbol{x}')=-\frac{\delta\Big(t-t'-\frac{|\boldsymbol{x}
-\boldsymbol{x}'|}{c}\Big)}{4\pi|\boldsymbol{x}-\boldsymbol{x}'|}
+\frac{m}{4\pi}\frac{J_{1}\left(mc\sqrt{(t-t')^{2}-\frac{|\boldsymbol{x}-\boldsymbol{x}'|^{2}}{c^2}}\right)}{\sqrt{(t-t')^{2}
-\frac{|\boldsymbol{x}-\boldsymbol{x}'|^{2}}{c^2}}}
H\left(t-t'-\frac{|\boldsymbol{x}-\boldsymbol{x}'|}{c}\right),%\tag{4.5}
\end{align}
where $J_{1}$ is the Bessel function of the first order and $H$ is the Heaviside's step function.  Thus,
\begin{align}
R^{(1)}(t,\boldsymbol{x})&=\int d^{3}x'\int dt'\mathcal{G}(t,\boldsymbol{x};t',\boldsymbol{x}')m^{2}
\kappa T^{(1)}(t',\boldsymbol{x}')\notag\\
\label{equ4.6}&=-\frac{m^2\kappa}{4\pi}\int d^{3}x'\frac{T\Big(t-\frac{|\boldsymbol{x}-\boldsymbol{x}'|}{c},\boldsymbol{x}'\Big)}{|\boldsymbol{x}-\boldsymbol{x}'|}
+\frac{m^3\kappa}{4\pi}\int d^{3}x'\int_{-\infty}^{t-\frac{|\boldsymbol{x}-\boldsymbol{x}'|}{c}} dt'
\frac{J_{1}\left(mc\sqrt{(t-t')^{2}-\frac{|\boldsymbol{x}-\boldsymbol{x}'|^{2}}{c^2}}\right)}{\sqrt{(t-t')^{2}
-\frac{|\boldsymbol{x}-\boldsymbol{x}'|^{2}}{c^2}}}T^{(1)}(t',\boldsymbol{x}').%\tag{4.6}
\end{align}
For convenience, let
\begin{align}
\label{equ4.7}\mathcal{G}(t,\boldsymbol{x};t',\boldsymbol{x}')&=\mathcal{G}_{1}(t,\boldsymbol{x};t',\boldsymbol{x}')
+\mathcal{G}_{2}(t,\boldsymbol{x};t',\boldsymbol{x}'),\\%\tag{4.7a}\\
\label{equ4.8}\mathcal{G}_{1}(t,\boldsymbol{x};t',\boldsymbol{x}'):&=-\frac{\delta\Big(t-t'-\frac{|\boldsymbol{x}-\boldsymbol{x}'|}{c}\Big)}{4\pi|\boldsymbol{x}-\boldsymbol{x}'|},\\%\tag{4.7b}\\
\label{equ4.9}\mathcal{G}_{2}(t,\boldsymbol{x};t',\boldsymbol{x}'):&=\frac{m}{4\pi}\frac{J_{1}\left(mc\sqrt{(t-t')^{2}-\frac{|\boldsymbol{x}-\boldsymbol{x}'|^{2}}{c^2}}\right)}{\sqrt{(t-t')^{2}-\frac{|\boldsymbol{x}-\boldsymbol{x}'|^{2}}{c^2}}}
H\left(t-t'-\frac{|\boldsymbol{x}-\boldsymbol{x}'|}{c}\right).%\tag{4.7c}
\end{align}
Correspondingly,
\begin{align}
\label{equ4.10}R^{(1)}(t,\boldsymbol{x})&=R^{(1)}_{1}(t,\boldsymbol{x})+R^{(1)}_{2}(t,\boldsymbol{x}),\\%\tag{4.8a}\\
\label{equ4.11}R^{(1)}_{1}(t,\boldsymbol{x}):&=\int d^{3}x'\int dt'\mathcal{G}_{1}(t,\boldsymbol{x};t',\boldsymbol{x}')m^{2}\kappa T^{(1)}(t',\boldsymbol{x}'),\\%\tag{4.8b}\\
\label{equ4.12}R^{(1)}_{2}(t,\boldsymbol{x}):&=\int d^{3}x'\int dt'\mathcal{G}_{2}(t,\boldsymbol{x};t',\boldsymbol{x}')m^{2}\kappa T^{(1)}(t',\boldsymbol{x}').%\tag{4.8c}
\end{align}
\end{widetext}

%%%%%%%%%%%%%     %%%%%%%%%%%%%%%
\subsubsection{The multipole expansion of $R^{(1)}_{1}$}

Dealing with $\mathcal{G}_{1}$ and $R^{(1)}_{1}$ is easy by Ref.~\cite{Campbell:1977jf}.  %Firstly, we should
Define an auxiliary variable:
\begin{align}
\nu_{1}:&=\frac{r^2+r'^2-c^2(t-t')^2}{2rr'},\notag\\%\tag{4.9a}\\
\label{equ4.13}&=\frac{r_{>}^2+r_{<}^2-c^2(t-t')^2}{2r_{>}r_{<}},%\tag{4.9b}
\end{align}
where $r=|\boldsymbol{x}|$, $r'=|\boldsymbol{x}'|$,
$r_{<}$ represents the lesser of $r$ and $r'$, and $r_{>}$ the greater.  Since
\begin{align}
\label{equ4.14}|\boldsymbol{x}-\boldsymbol{x}'|^{2}&=r^2+r'^2-2rr'\cos{\tilde{\theta}}\notag\\
&=r_{>}^2+r_{<}^2-2r_{>}r_{<}\cos{\tilde{\theta}},%\tag{4.10}
\end{align}
\begin{align}
\label{equ4.15}\cos{\tilde{\theta}}-\nu_{1}=\frac{c^2(t-t')^{2}-|\boldsymbol{x}-\boldsymbol{x}'|^{2}}{2r_{>}r_{<}},%\tag{4.11}
\end{align}
where $\cos{\tilde{\theta}}=\boldsymbol{n}'\cdot\boldsymbol{n}$ as in Sec.~\ref{Sec:Preliminary}, and $\boldsymbol{n},\boldsymbol{n}'$ are the unit vectors of $\boldsymbol{x},\boldsymbol{x}'$, respectively. Eq.~(\ref{equ4.15}) in turn leads to
 \begin{align*}
\label{equ4.16} -1 \leq \frac{c^2(t-t')^{2}-|\boldsymbol{x}-\boldsymbol{x}'|^{2}}{2r_{>}r_{<}}+\nu_{1} =\cos{\tilde{\theta}}\leq 1.
\end{align*}
Because of the presence of the delta function in (\ref{equ4.8}),
 \begin{align}
-1\leq \nu_{1}=\cos{\tilde{\theta}}\leq 1.
\end{align}
With the help of (\ref{equ4.15}) and the two properties of the %\pagebreak
Dirac delta function,
\begin{align}
\label{equ4.17}\delta{(ax)}&=\frac{1}{|a|}\delta{(x)},\\%\tag{4.13a}\\
\label{equ4.18}\delta{(x^{2}-a^{2})}&=\frac{1}{2|a|}\big(\delta(x+a)+\delta(x-a)\big),%\tag{4.13b}
\end{align}
the retarded Green's function is related to $\cos{\tilde{\theta}}-\nu_{1}$ by

\begin{align}
\label{equ4.19}\delta(\cos{\tilde{\theta}}-\nu_{1})H(t-t')& =-\frac{4\pi r_{>}r_{<}}{c}\mathcal{G}_{1}, %\tag{4.14a}
\end{align}
or %\pagebreak
\begin{align}
\label{equ4.20}\mathcal{G}_{1}
=-\frac{c}{4\pi r_{>}r_{<}}\delta(\cos{\tilde{\theta}}-\nu_{1})H(t-t')H(1-|\nu_{1}|),%\tag{4.14b}
\end{align}
where the addition of $H(1-|\nu_{1}|)$ in the latter equation will, at most, affect the boundary values of ${\cal G}_1$ at $\nu_1=\pm1$, which
will not affect the value of $R_1^{(1)}$.  Again, with the help of the property of the Dirac delta function~\cite{olver2010nist},
\begin{align}
\label{equ4.21}\delta{(y-x)}&=\frac{1}{2}\sum_{l=0}^{\infty}(2l+1)P_{l}(y)P_{l}(x)%\tag{4.15}
\end{align}
and Eq.~(\ref{equ2.17}), the retarded Green's function (\ref{equ4.20}) can be rewritten as
\begin{widetext}
\begin{align}
\mathcal{G}_{1}
\label{equ4.22}&=-\frac{c}{8\pi r_{>}r_{<}}H(t-t')H(1-|\nu_{1}|)\sum_{l=0}^{\infty}\frac{(2l+1)!!}{l!}P_{l}(\nu_{1})\hat{N}_{I_{l}}(\theta',\varphi')\hat{N}_{I_{l}}(\theta,\varphi),%\tag{4.16}
\end{align}
\end{widetext}
where $(\theta',\varphi')$ and $(\theta,\varphi)$ are the angle coordinates of $\boldsymbol{x}'$ and $\boldsymbol{x}$, respectively.

In order to derive the multipole expansion of $R^{(1)}_{1}$, the above expression of $\mathcal{G}_{1}$ should be simplified.  Define
a new variable
\begin{align}
\label{equ4.23}z_{1}&:=\frac{c(t'-t)+r_{>}}{r_{<}}%\tag{4.17}
\end{align}
to replace $t'$.  Obviously, $z_1$ satisfies
\begin{align}
\label{equ4.24}dz_{1}&=\frac{c}{r_{<}}dt',\\%\tag{4.18a}\\
\label{equ4.25}t'&=t-\frac{r_{>}}{c}+\frac{r_{<}z_{1}}{c}.%\tag{4.18b}
\end{align}
In addition,
\begin{align}
\label{equ4.26}|z_{1}|&\leqslant 1,\\%\tag{4.18c}\\
\label{equ4.27}\nu_{1}&=z_{1}+\frac{r_{<}}{2r_{>}}(1-z_{1}^2).%\tag{4.18d}
\end{align}
From (\ref{equ4.13}), the retarded property ($t>t'$), (\ref{equ4.16}), and (\ref{equ4.23}), the inequality (\ref{equ4.26})
can be easily proved. Eq. (\ref{equ4.27}) is the direct result of the definitions of $z_{1}$ and $\nu_{1}$.
Now, we will simplify (\ref{equ4.22}) by use of the above formulas. The Taylor expansion of $P_{l}(\nu_{1})$ around
$\nu_{1}=z_{1}$ is
\begin{align*}
P_{l}(\nu_{1})&=\sum_{j=0}^{\infty} \left . \frac{1}{j!}\frac{d^{j}P_{l}(\nu_{1})}{d\nu_{1}^{j}}\right |_{\nu_{1}=z_{1}}(\nu_{1}-z_{1})^{j}.
\end{align*}
Since $P_{l}(\nu_{1})$ is a polynomial of degree $l$, the Taylor expansion reduces to
\begin{align}
\label{equ4.28}P_{l}(\nu_{1})&=\sum_{j=0}^{l}\frac{1}{2^jj!}\Big(\frac{r_{<}}{r_{>}}\Big)^{j}(1-z_{1}^2)^{^{j/2}}P_{l}^{j}(z_{1})%\tag{4.19}
\end{align}
on account of (\ref{equ4.27}), where
\begin{align}
\label{equ4.29}P_{l}^{j}(z_{1})=(1-z_{1}^2)^{j/2}\frac{d^{j}P_{l}(z_{1})}{dz_{1}^{j}}%\tag{4.20}
\end{align}
is the associated Legendre polynomial.
Inserting (\ref{equ4.28}) into (\ref{equ4.22}) gives
\begin{widetext}
\begin{align}
\label{equ4.30}\mathcal{G}_{1}=-\frac{c}{8\pi r_{>}r_{<}}H(t-t')H(1-|z_{1}|)\sum_{l=0}^{\infty}\frac{(2l+1)!!}{l!}\sum_{j=0}^{l}\frac{1}{2^jj!}\left ( \frac{r_{<}}{r_{>}}\right )^{j}(1-z_{1}^2)^{j/2}P_{l}^{j}(z_{1})\hat{N}_{I_{l}}(\theta',\varphi')\hat{N}_{I_{l}}(\theta,\varphi).%\tag{4.21}
\end{align}
This is the expression of $\mathcal{G}_{1}$ to derive the multipole expansion of $R^{(1)}_{1}$.
By using Eqs.~(\ref{equ4.11}), (\ref{equ4.24})---(\ref{equ4.26}), and (\ref{equ4.30}), we obtain
\begin{align}
R^{(1)}_{1}(t,\boldsymbol{x})=-\frac{m^{2}\kappa}{8\pi r_{>}}\int d^{3}x'\int_{-1}^{1}dz_{1}
\sum_{l=0}^{\infty}\frac{(2l+1)!!}{l!}\sum_{j=0}^{l}\frac{1}{2^jj!}\left(\frac{r_{<}}{r_{>}}\right)^{j}(1-z_{1}^2)^{j/2}P_{l}^{j}(z_{1})
\hat{N}_{I_{l}}(\theta',\varphi')\hat{N}_{I_{l}}(\theta,\varphi)\times \notag\\
\label{equ4.31}T^{(1)}\Big(t-\frac{r_{>}}{c}+\frac{r_{<}z_{1}}{c},\boldsymbol{x}'\Big).%\tag{4.22}
\end{align}
We consider the points outside the source region, namely $r=r_{>}$ and $r'=r_{<}$.  Substituting the equality~\cite{Campbell:1977jf}
\begin{align}
\label{equ4.32}&(1-z_{1}^2)^{j/2}P_{l}^{j}(z_{1})=\frac{(-1)^{l-j}}{2^ll!}\frac{(l+j)!}{(l-j)!}\frac{d^{l-j}}{dz_{1}^{l-j}}(1-z_{1}^2)^l%\tag{4.23}
\end{align}
into (\ref{equ4.31}) and then making use of (\ref{equ4.24})---(\ref{equ4.26}) again, Eq.~(\ref{equ4.31}) can be rewritten in the form in terms of
$t'$-integration,
\begin{align}
R^{(1)}_{1}(t,\boldsymbol{x})=-\frac{m^{2}\kappa c}{4\pi}\sum_{l=0}^{\infty}\frac{(-1)^l}{l!}\int d^{3}x'\int_{u-\frac{r'}{c}}^{u+\frac{r'}{c}}dt'
\hat{N}_{I_{l}}(\theta,\varphi)\sum_{j=0}^{l}\frac{(-1)^j}{2^jj!}\frac{(l+j)!}{(l-j)!}\frac{1}{c^{l-j}r^{j+1}}\times\notag\\
\label{equ4.33}\left[\frac{d^{l-j}}{dt'^{l-j}}\Big(1-\frac{c^2}{r'^2}(t'-u)^2\Big)^l\right]\frac{(2l+1)!!}{2^{l+1}l!}r'^{l-1}
\hat{N}_{I_{l}}(\theta',\varphi')
T^{(1)}(t',\boldsymbol{x}').%\tag{4.22}
\end{align}
The derivative $\dfrac{d}{dt'}$ in the above expression can be replaced by $-\dfrac{d}{du}$.  Then, %we acquire
\begin{align}
\label{equ4.34}R^{(1)}_{1}(t,\boldsymbol{x})=-\frac{m^{2}\kappa c}{4\pi}\sum_{l=0}^{\infty}\frac{(-1)^l}{l!}\int d^{3}x'\int_{u-\frac{r'}{c}}^{u+\frac{r'}{c}}dt'
\hat{N}_{I_{l}}(\theta,\varphi)\sum_{j=0}^{l}\frac{(-1)^l}{2^jj!}\frac{(l+j)!}{(l-j)!}\frac{1}{c^{l-j}r^{j+1}}\times\notag\\
\left[\frac{d^{l-j}}{du^{l-j}}\Big(1-\frac{c^2}{r'^2}(t'-u)^2\Big)^l\right]\frac{(2l+1)!!}{2^{l+1}l!}r'^{l-1}
\hat{N}_{I_{l}}(\theta',\varphi')
T^{(1)}(t',\boldsymbol{x}').%\tag{4.22}
\end{align}
Further, inserting (\ref{equ2.14}) with $\epsilon=1$ into (\ref{equ4.34})
gives the multipole expansion of $R^{(1)}_{1}$
\begin{align}
\label{equ4.35}R^{(1)}_{1}(t,\boldsymbol{x})=-\frac{m^{2}\kappa c}{4\pi}\sum_{l=0}^{\infty}\frac{(-1)^l}{l!}\int d^{3}x'\int_{u-\frac{r'}{c}}^{u+\frac{r'}{c}}dt'
\frac{1}{r'}\hat{X'}_{I_{l}}(\theta',\varphi')\hat{\mathcal{T}}_{1I_{l}}(u;t',\boldsymbol{x}'),
\end{align}
where
\begin{align}
\label{equ4.36}
X'_{I_{l}}(\theta',\varphi')=X'_{i_{1}i_{2}\cdots i_{l}}(\theta',\varphi'):= x'_{i_{1}}x'_{i_{2}}\cdots x'_{i_{l}}={r'}^l N_{I_l}(\theta',\varphi')%\tag{4.29}
\end{align}
and
\begin{align}
\label{equ4.37}\hat{\mathcal{T}}_{1I_{l}}(u;t',\boldsymbol{x}'):=\frac{(2l+1)!!}{2^{l+1}l!}\hat{\partial}_{I_{l}}\left(\frac{1}{r}\Big(1-\frac{c^2}{r'^2}(t'-u)^2\Big)^l\right)
T^{(1)}(t',\boldsymbol{x}').%\tag{4.31}
\end{align}

In fact, we can derive the traditional form of (\ref{equ4.35}) in a similar way in
Ref.~\cite{Blanchet:2013haa,Damour:1990gj} from (\ref{equ4.31}) and (\ref{equ4.32}), namely
in terms of $z_{1}$-integration
\begin{align}
\label{equ4.38}R^{(1)}_{1}(t,\boldsymbol{x})
&=-\frac{m^{2}\kappa}{4\pi}\sum_{l=0}^{\infty}\frac{(-1)^l}{l!}\partial_{I_{l}}\Big(\frac{\hat{F}_{I_{l}}(u)}{r}\Big),%\tag{4.27a}
\end{align}
where
\begin{align}
\label{equ4.39}
\hat{F}_{I_{l}}(u)&=\int d^{3}x'
{\hat X}'_{I_{l}}(\theta',\varphi')
\overline{T}_{l}(u,\boldsymbol{x}')%\tag{4.27b}
\end{align}
is the $l$-pole moment, and
\begin{align}
\label{equ4.40}\overline{T}_{l}(u,\boldsymbol{x}'):=\frac{(2l+1)!!}{2^{l+1}l!}\int_{-1}^{1}(1-z_{1}^2)^l
T^{(1)}(u+\frac{r'z_{1}}{c},\boldsymbol{x}')dz_{1}.%\tag{4.28}
\end{align}

\subsubsection{The multipole expansion of $R^{(1)}_{2}$}

Dealing with $\mathcal{G}_{2}$ and $R^{(1)}_{2}$ is not easy like $\mathcal{G}_{1}$ and $R^{(1)}_{1}$.
The retarded Green's function (\ref{equ4.9})
can be written as
\begin{align}
\label{equ4.41}\mathcal{G}_{2}=&\frac{m^2c}{4\pi}H(t-t')\frac{J_{1}\left(m\sqrt{c^2(t-t')^{2}-|\boldsymbol{x}-\boldsymbol{x}'|^{2}}\right)}{m\sqrt{c^2(t-t')^{2}-|\boldsymbol{x}-\boldsymbol{x}'|^{2}}}
H\big(c^2(t-t')^2-|\boldsymbol{x}-\boldsymbol{x}'|^2\big).%\tag{4.33}
\end{align}
As in the previous subsubsection, introduce an auxiliary variable
\begin{align}
\label{equ4.42}\nu_{2}:&=\frac{r^2+r'^2-c^2(t-t')^2}{2rr'},%\tag{4.36a}\\
\end{align}
with which
\begin{align}
\label{equ4.43}c^2(t-t')^{2}-|\boldsymbol{x}-\boldsymbol{x}'|^{2}=2rr'(\cos{\tilde{\theta}}-\nu_{2}).%\tag{4.37}
\end{align}
It will be seen very soon that the range of $\nu_2$ is different from that of $\nu_1$ although their definitions are the
same.  Then,
\begin{align}
\label{equ4.44}
\mathcal{G}_{2}=&\frac{m^2c}{4\pi}H(t-t')\frac{J_{1}\left(m\sqrt{2rr'(\cos{\tilde{\theta}}-\nu_{2})}\right)}{m\sqrt{2rr'(\cos{\tilde{\theta}}-\nu_{2})}}
H(\cos{\tilde{\theta}}-\nu_{2}).%\tag{4.33}
\end{align}
\end{widetext}
The definition of Heaviside's step function shows
\begin{align}
\label{equ4.45}\nu_{2}\leq\cos{\tilde{\theta}} \in [-1,1].%\tag{4.39}
\end{align}
Remember that $(t',\boldsymbol{x}')$ is the spacetime point of a source and $(t,\boldsymbol{x})$ is the spacetime point of observation.
The detection of GWs is usually made in the region
\begin{align}
\label{equ4.46}\frac{r'}{r}\lll 1.%\tag{4.40}
\end{align}
In this region, the leading term of $\nu_2$ is independent of $\tilde \theta$.  Therefore, the leading term of $\nu_2$ should satisfy
\begin{align}
\label{equ4.47}\nu_{2}\leq-1,%\tag{4.42}
\end{align}
if we focus on the fields at the observation point.  Then, (\ref{equ4.44}) can be approximated very well by
\begin{align}
\label{equ4.48}
\mathcal{G}_{2}=&\frac{m^2c}{4\pi}H(t-t')H(-1-\nu_{2})\frac{J_{1}\left(m\sqrt{2rr'(\cos{\tilde{\theta}}-\nu_{2})}\right)}{m\sqrt{2rr'(\cos{\tilde{\theta}}-\nu_{2})}}
.%\tag{4.33}
\end{align}

From (\ref{equ4.48}), we see that $\mathcal{G}_{2}$ is a function of $\cos{\tilde{\theta}}$.  For the emphasis of the dependence on $\cos\tilde \theta$, define

\begin{align}
\label{equ4.49}\mathcal{K}(\cos{\tilde{\theta}}):=\frac{J_{1}\left(\sqrt{2m^2rr'(\cos{\tilde{\theta}}-\nu_{2})}\right)}
{\sqrt{2m^2rr'(\cos{\tilde{\theta}}-\nu_{2})}}.%\tag{4.34}
\end{align}
Then,
\begin{align}
\label{equ4.50}\mathcal{G}_{2}=&\frac{m^2c}{4\pi}H(t-t')H(-1-\nu_{2})\mathcal{K}(\cos{\tilde{\theta}}).%\tag{4.35}
\end{align}
Note that the Legendre polynomials form a complete set of functions in the range $|\cos{\tilde{\theta}}|\leq 1$. $\mathcal{K}(\cos{\tilde{\theta}})$ can be expanded in terms of the Legendre polynomial $P_{l}(\cos{\tilde{\theta}})$ as
\begin{align}
\label{equ4.51}&\mathcal{K}(\cos{\tilde{\theta}})=\sum_{l=0}^{\infty}c_{l}P_{l}(\cos{\tilde{\theta}}),%\tag{4.43a}
\end{align}
where
\begin{align}
\label{equ4.52}&c_{l}=\frac{2l+1}{2}\int_{0}^{\pi}\mathcal{K}(\cos{\tilde{\theta}})P_{l}(\cos{\tilde{\theta}})\sin{\tilde{\theta}}d\tilde{\theta}.%\tag{4.43b}
\end{align}
The detailed calculation of the coefficient $c_{l}$ is tedious and is put in Appendix A.  The result is
\begin{align}
\label{equ4.53}&\qquad\qquad c_{l}=\frac{2l+1}{2m^2r r' }\mathcal{P}_{l}(\nu_{2}),%\tag{4.44a}
\end{align}
where
\begin{widetext}
\begin{align}
\label{equ4.54}\mathcal{P}_{l}(\nu_{2})=&
\left [\exp{\left ( \frac{m^2r r' }{2}\left(\frac{d}{d\nu_{2}}\right)_{-}^{-1}\right)}-\exp{\left (\frac{m^2r r' }{2}\left(\frac{d}{d\nu_{2}}\right)_{+}^{-1}\right )}\right ]P_{l}(\nu_{2}).%\tag{4.44b}
\end{align}
In (\ref{equ4.54}), the range of the argument of the Legendre polynomial has been extended to $\nu_{2}\leq-1$, and the operators
$\left(\dfrac{d}{d\nu_{2}}\right)_{\pm}^{-1}$
for an arbitrary function $\tilde{f}(\nu_{2})$ is defined by
\begin{align}
\label{equ4.55}\left(\frac{d}{d\nu_{2}}\right)_{\pm}^{-1}\tilde{f}(\nu_{2}):&=-\int_{\nu_{2}}^{\pm1}\tilde{f}(s)ds.
%\qquad \left(\frac{d}{d\nu_{2}}\right)_{-}^{-1}\tilde{f}(\nu_{2}):=-\int_{\nu_{2}}^{-1}\tilde{f}(s)ds.%\tag{4.45}
\end{align}
Together with (\ref{equ2.17}), (\ref{equ4.51}), and (\ref{equ4.53}), Eq.~(\ref{equ4.50})
can be written as
\begin{align}
\label{equ4.56}\mathcal{G}_{2}
&=\frac{c}{8\pi r r' }H(t-t')H(-1-\nu_{2})\sum_{l=0}^{\infty}\frac{(2l+1)!!}{l!}\mathcal{P}_{l}(\nu_{2})\hat{N}_{I_{l}}(\theta',\varphi')\hat{N}_{I_{l}}(\theta,\varphi),%\tag{4.46}
\end{align}
\end{widetext}
where the meanings of $(\theta',\varphi')$ and $(\theta,\varphi)$ are the same as before.

In order to derive the multipole expansion of $R^{(1)}_{2}$,  define a new variable again
\begin{align}
\label{equ4.57}z_{2}&:=\frac{c(t'-t)+r }{r' }%\tag{4.47}
\end{align}
to replace $t'$.  The definition of $z_2$ is the same as $z_{1}$, but they take values in different ranges.  $z_2$ satisfies
\begin{align}
\label{equ4.58}dz_{2}&=\frac{c}{r' }dt',\\%\tag{4.48a}\\
%\end{align*}
%\begin{align*}
\label{equ4.59}t'&=t-\frac{r }{c}+\frac{r' z_{2}}{c},\\%\tag{4.48b}
\label{equ4.60}z_{2}&\leq-1,%\tag{4.48c} \\
\end{align}
\pagebreak
\begin{align}
\label{equ4.61}\nu_{2}&=z_{2}+\frac{r' }{2r }(1-z_{2}^2).%\tag{4.48d}
\end{align}
Their proofs are also similar to those of $z_{1}$.

We can also make a Taylor expansion of $P_{l}(\nu_{2})$ around $\nu_{2}=z_{2}$ to simplify (\ref{equ4.56}). Similar to
(\ref{equ4.28}), we have
\begin{align}
\label{equ4.62}P_{l}(\nu_{2})&=\sum_{j=0}^{l}\frac{1}{2^jj!}\Big(\frac{r' }{r }\Big)^{j}(1-z_{2}^2)^{j/2}P_{l}^{j}(z_{2}),%\tag{4.49}
\end{align}
where $P_{l}^{j}$ is again an associated Legendre polynomial. Even for odd $j$,
$(1-z_{2}^2)^{j/2}P_{l}^{j}(z_{2})$ remains real though both $(1-z_{2}^2)^{j/2}$ and $P_{l}^{j}(z_{2})$ are imaginary.
Inserting (\ref{equ4.54}) and (\ref{equ4.62}) into (\ref{equ4.56}) gives
\begin{widetext}
\begin{align}
\mathcal{G}_{2}=\frac{c}{8\pi r r' }H(t-t')H(-1-z_{2})\sum_{l=0}^{\infty}&\frac{(2l+1)!!}{l!}
\left(\exp{\Big(\frac{m^2r r' }{2}(\text{int}\ z_{2})_{-}\Big)}-\exp{\Big(\frac{m^2r r' }{2}(\text{int}\ z_{2})_{+}\Big)}\right)\notag\\
\label{equ4.63}&\sum_{j=0}^{l}\frac{1}{2^jj!}\Big(\frac{r' }{r }\Big) ^{j} (1-z_{2}^2)^{j/2}P_{l}^{j}(z_{2})\hat{N}_{I_{l}}(\theta',\varphi')\hat{N}_{I_{l}}(\theta,\varphi),%\tag{4.50}
\end{align}
where the operators $\left(\dfrac{d}{d\nu_{2}}\right)_{\pm}^{-1}$ %and $\big(\frac{d}{d\nu_{2}}\big)_{-}^{-1}$
have been changed to $(\text{int}\ z_{2})_{\pm}$ %and $(\text{int}\ z_{2})_{-}$
by replacing the variable $\nu_{2}$ with $z_{2}$, namely for an arbitrary function $\tilde{f}(z_{2})$
\begin{align}
\label{equ4.64}(\text{int}\ z_{2})_{\pm}\tilde{f}(z_{2}):&=-\int_{z_{2}}^{\pm 1}\tilde{f}(q)\Big(1-\frac{r' }{r }q\Big)dq.
%\qquad (\text{int}\ z_{2})_{-}\tilde{f}(z_{2}):=-\int_{z_{2}}^{-1}\tilde{f}(q)(1-\frac{r' }{r }q)dq.\tag{4.51}
\end{align}
Eq.~(\ref{equ4.63}) is the expression of $\mathcal{G}_{2}$ to obtain the multipole expansion of $R^{(1)}_{2}$.
By (\ref{equ4.12}), (\ref{equ4.58})---(\ref{equ4.60}), and (\ref{equ4.63}), we can acquire
\begin{align}
\label{equ4.65}
R^{(1)}_{2}(t,\boldsymbol{x})&=\frac{m^{2}\kappa}{8\pi r }\int d^{3}x'\int_{-\infty}^{-1}dz_{2}
\sum_{l=0}^{\infty}\frac{(2l+1)!!}{l!}\sum_{j=0}^{l}\frac{1}{2^jj!}\Big(\frac{r'}{r}\Big)^{j}\hat{N}_{I_{l}}(\theta',\varphi')\hat{N}_{I_{l}}(\theta,\varphi)\times\notag\\
&\qquad\left[\left(\exp{\Big(\frac{m^2r r' }{2}(\text{int}\ z_{2})_{-}\Big)}-\exp{\Big(\frac{m^2r r' }{2}(\text{int}\ z_{2})_{+}\Big)}\right)(1-z_{2}^2)^{j/2}P_{l}^{j}(z_{2})\right]
T^{(1)}(u+\frac{r' z_{2}}{c},\boldsymbol{x}').%\tag{4.52}
\end{align}
%We consider the points outside the source region, namely $r=r $ and $r'=r' $, and
Plugging equality~\cite{Campbell:1977jf}
\begin{align}
\label{equ4.66}&(1-z_{2}^2)^{j/2}P_{l}^{j}(z_{2})=\frac{(-1)^{l-j}}{2^ll!}\frac{(l+j)!}{(l-j)!}\frac{d^{l-j}}{dz_{2}^{l-j}}(1-z_{2}^2)^l%\tag{4.53}
\end{align}
into (\ref{equ4.65}) gives
\begin{align}
\label{equ4.67}
R^{(1)}_{2}(t,\boldsymbol{x})=&\frac{m^{2}\kappa}{4\pi}\sum_{l=0}^{\infty}\frac{(-1)^l}{l!}\int d^{3}x'\int_{-\infty}^{-1}dz_{2}\hat{N}_{I_{l}}(\theta,\varphi)\sum_{j=0}^{l}\frac{(-1)^{j}}{2^jj!}\frac{(l+j)!}{(l-j)!}\frac{1}{r^{j+1}}
\frac{(2l+1)!!}{2^{l+1}l!}r'^{j}\hat{N}_{I_{l}}(\theta',\varphi')\times\notag\\
&\left[\left(\exp{\Big(\frac{m^2rr'}{2}(\text{int}\ z_{2})_{-}\Big)}
-\exp{\Big(\frac{m^2rr'}{2}(\text{int}\ z_{2})_{+}\Big)}\right)\frac{d^{l-j}}{dz_{2}^{l-j}}(1-z_{2}^2)^l\right]
T^{(1)}\Big(u+\frac{r'z_{2}}{c},\boldsymbol{x}'\Big).%\tag{4.54}
\end{align}
By use of (\ref{equ4.58})---(\ref{equ4.60}) again, the above $R^{(1)}_{2}$ can be transformed into the expression in terms of $t'$-integral,
\begin{align}
\label{equ4.68}R^{(1)}_{2}(t,\boldsymbol{x})
=&\frac{m^{2}\kappa c}{4\pi}\sum_{l=0}^{\infty}\frac{(-1)^l}{l!}\int d^{3}x'\int_{-\infty}^{u-\frac{r'}{c}}dt'\hat{N}_{I_{l}}(\theta,\varphi)\sum_{j=0}^{l}\frac{(-1)^{j}}{2^jj!}\frac{(l+j)!}{(l-j)!}
\frac{1}{c^{l-j}r^{j+1}}\frac{(2l+1)!!}{2^{l+1}l!}r'^{l-1}\hat{N}_{I_{l}}(\theta',\varphi')\times\notag\\
&\left[\left(\exp{\Big(\frac{m^2rr'}{2}(\text{int}\ t')_{-}\Big)}
-\exp{\Big(\frac{m^2rr'}{2}(\text{int}\ t')_{+}\Big)}\right)\frac{d^{l-j}}{dt'^{l-j}}\Big(1-\frac{c^2}{r'^2}(t'-u)^2\Big)^l\right]
T^{(1)}(t',\boldsymbol{x}'),%\tag{4.56}
\end{align}
where the operators $(\text{int}\ t')_{\pm}$ for an arbitrary function $\tilde{f}(t')$ are defined as
\begin{align}
\label{equ4.69}(\text{int}\ t')_{\pm}\tilde{f}(t'):&=-\frac{c}{r'}\int_{t'}^{u\pm\frac{r'}{c}}\tilde{f}(\tau)\Big(1+\frac{c}{r}(u-\tau)\Big)d\tau.
\end{align}
In (\ref{equ4.68}), $\dfrac{d}{dt'}$ can be replaced by  $-\dfrac{d}{du}$, so
\begin{align}
\label{equ4.70}R^{(1)}_{2}(t,\boldsymbol{x})
=&\frac{m^{2}\kappa c}{4\pi}\sum_{l=0}^{\infty}\frac{(-1)^l}{l!}\int d^{3}x'\int_{-\infty}^{u-\frac{r'}{c}}dt'
\bigg[\left(\exp{\Big(\frac{m^2rr'}{2}(\text{int}\ t')_{-}\Big)}
-\exp{\Big(\frac{m^2rr'}{2}(\text{int}\ t')_{+}\Big)}\right)\notag\\
&\hat{N}_{I_{l}}(\theta,\varphi)\sum_{j=0}^{l}\frac{(-1)^{l}}{2^jj!}\frac{(l+j)!}{(l-j)!}\frac{1}{c^{l-j}r^{j+1}}
\frac{d^{l-j}}{du^{l-j}}\Big(1-\frac{c^2}{r'^2}(t'-u)^2\Big)^l\bigg]
\frac{(2l+1)!!}{2^{l+1}l!}
\frac{\hat{X'}_{I_{l}}(\theta',\varphi')}{r'}T^{(1)}(t',\boldsymbol{x}'),%\tag{4.58}
\end{align}
where the meaning of $\hat{X'}_{I_{l}}(\theta',\varphi')$ is the same as (\ref{equ4.36}).  With the help of Eq.~(\ref{equ2.14}) with $\epsilon=1$, we obtain the multipole expansion of $R^{(1)}_{2}$:
\begin{align}
\label{equ4.71}R^{(1)}_{2}(t,\boldsymbol{x})=\frac{m^{2}\kappa c}{4\pi}\sum_{l=0}^{\infty}\frac{(-1)^l}{l!}\int d^{3}x'\int_{-\infty}^{u-\frac{r'}{c}}dt'
\frac{1}{r'}\hat{X'}_{I_{l}}(\theta',\varphi')\hat{\mathcal{T}}_{2I_{l}}(t,r;t',\boldsymbol{x}'),
\end{align}
where
\begin{align}
\label{equ4.72}\hat{\mathcal{T}}_{2I_{l}}(u,r;t',\boldsymbol{x}'):=\frac{(2l+1)!!}{2^{l+1}l!}
\left[\left(\exp{\Big(\frac{m^2rr'}{2}(\text{int}\ t')_{-}\Big)}
-\exp{\Big(\frac{m^2rr'}{2}(\text{int}\ t')_{+}\Big)}\right)\hat{\partial}_{I_{l}}\left(\frac{1}{r}\Big(1-\frac{c^2}{r'^2}(t'-u)^2\Big)^l\right)\right]
T^{(1)}(t',\boldsymbol{x}').%\tag{4.31}
\end{align}

Upon obtaining (\ref{equ4.35}) and (\ref{equ4.71}), we can write down the multipole expansion of $R^{(1)}$ by (\ref{equ4.10}).  It is
\begin{align}
\label{equ4.73}R^{(1)}(t,\boldsymbol{x})
&=-\frac{m^{2}\kappa c}{4\pi}\sum_{l=0}^{\infty}\frac{(-1)^l}{l!}\int d^{3}x'\frac{1}{r'}\hat{X'}_{I_{l}}(\theta',\varphi')\left(\int_{u-\frac{r'}{c}}^{u+\frac{r'}{c}}dt'\hat{\mathcal{T}}_{1I_{l}}(u;t',\boldsymbol{x}')
-\int_{-\infty}^{u-\frac{r'}{c}}dt'\hat{\mathcal{T}}_{2I_{l}}(t,r;t',\boldsymbol{x}')\right).%\tag{4.60}
\end{align}

%%%%%%%%%%%%%%     expansion of h^{\mu\nu}        %%%%%%%%%%%%%%%%%

\subsection{The multipole expansion of $h^{\mu\nu}$}

We have known that the multipole expansion of $h^{\mu\nu}$ contains two parts: the tensor part
associated with $\tilde{h}^{\mu\nu}$ and the scalar part associated with $R^{(1)}$. (\ref{equ4.1}) shows that
these two parts are $\tilde{h}^{\mu\nu}$ and $-2a\eta^{\mu\nu}R^{(1)}$, respectively.
The multipole expansion of $\tilde{h}^{\mu\nu}$ is given by (\ref{equ4.2}), and the multipole expansion of $-2a\eta^{\mu\nu}R^{(1)}$ at the
observation point is given by (\ref{equ3.64}), (\ref{equ4.46}), (\ref{equ4.73}) and $\kappa=8\pi G/c^4$, namely,
\begin{align}
\label{equ4.74}-2a\eta^{\mu\nu}R^{(1)}
&=\frac{2G}{3c^3}\eta^{\mu\nu}\sum_{l=0}^{\infty}\frac{(-1)^l}{l!}\int d^{3}x'\frac{1}{r'}\hat{X'}_{I_{l}}(\theta',\varphi')\left(\int_{u-\frac{r'}{c}}^{u+\frac{r'}{c}}dt'\hat{\mathcal{T}}_{1I_{l}}(u;t',\boldsymbol{x}')
-\int_{-\infty}^{u-\frac{r'}{c}}dt'\hat{\mathcal{T}}_{2I_{l}}(t,r;t',\boldsymbol{x}')\right),\quad \frac{r'}{r}\lll 1.%\tag{4.62}
\end{align}
Moreover, by (\ref{equ4.1}) and (\ref{equ4.2}), we derive the multipole expansion of $h^{\mu\nu}$ under the condition $r'/r\lll 1$:
\begin{equation}\label{equ4.75}
\left\{\begin{array}{l}
\displaystyle h^{00}(t,\boldsymbol{x})= -\frac{4G}{c^{2}}\sum_{l=0}^{\infty}\frac{(-1)^{l}}{l!}\partial_{I_{l}}\left(\frac{\hat{M}_{I_{l}}(u)}{r}\right)\\
\displaystyle \phantom{h^{00}(t,\boldsymbol{x})=}-\frac{2G}{3c^3}\sum_{l=0}^{\infty}\frac{(-1)^l}{l!}\int d^{3}x'\frac{1}{r'}\hat{X'}_{I_{l}}(\theta',\varphi')\left(\int_{u-\frac{r'}{c}}^{u+\frac{r'}{c}}dt'\hat{\mathcal{T}}_{1I_{l}}(u;t',\boldsymbol{x}')
-\int_{-\infty}^{u-\frac{r'}{c}}dt'\hat{\mathcal{T}}_{2I_{l}}(t,r;t',\boldsymbol{x}')\right),\\
\displaystyle h^{0i}(t,\boldsymbol{x})= \frac{4G}{c^{3}}\sum_{l=1}^{\infty}\frac{(-1)^{l}}{l!}\partial_{I_{l-1}}\left(\frac{\partial_{t}\hat{M}_{iI_{l-1}}(u)}{r}\right)+\frac{4G}{c^{3}}\sum_{l=1}^{\infty}\frac{(-1)^{l}l}{(l+1)!}\epsilon_{iab}\partial_{aI_{l-1}}\left(\frac{\hat{S}_{bI_{l-1}}(u)}{r}\right),\\
\displaystyle h^{ij}(t,\boldsymbol{x})=-\frac{4G}{c^{4}}\sum_{l=2}^{\infty}\frac{(-1)^{l}}{l!}\partial_{I_{l-2}}\left(\frac{\partial_{t}^{2}\hat{M}_{ijI_{l-2}}(u)}{r}\right)-\frac{8G}{c^{4}}\sum_{l=2}^{\infty}\frac{(-1)^{l}l}{(l+1)!}\partial_{aI_{l-2}}\left(\frac{\epsilon_{ab(i}\partial_{t}\hat{S}_{j)bI_{l-2}}(u)}{r}\right)\\
\displaystyle \phantom{h^{00}(t,\boldsymbol{x})=}+\frac{2G}{3c^3}\delta^{ij}\sum_{l=0}^{\infty}\frac{(-1)^l}{l!}\int d^{3}x'\frac{1}{r'}\hat{X'}_{I_{l}}(\theta',\varphi')\left(\int_{u-\frac{r'}{c}}^{u+\frac{r'}{c}}dt'\hat{\mathcal{T}}_{1I_{l}}(u;t',\boldsymbol{x}')
-\int_{-\infty}^{u-\frac{r'}{c}}dt'\hat{\mathcal{T}}_{2I_{l}}(t,r;t',\boldsymbol{x}')\right).
\end{array}\right.%\tag{4.63}
\end{equation}
\end{widetext}
This expression shows again that there exist monopole and dipole radiation for $f(R)$ gravity, which makes its prediction about GWs different from
the case in GR.

%%%%%%%%%%%%%%%%%%%%%%%%%%%%%%%%%%%%%%%%%%%%%%%%%%%%%%%
%%%%%%%%%%%%%%%%%%%%%%%%%%%%%%%%%%%%%%%%%%%%%%%%%%%%%%%
%%%           Stationary Cases                     %%%%
%%%%%%%%%%%%%%%%%%%%%%%%%%%%%%%%%%%%%%%%%%%%%%%%%%%%%%%
%%%%%%%%%%%%%%%%%%%%%%%%%%%%%%%%%%%%%%%%%%%%%%%%%%%%%%%

\section{\uppercase{The stationary multipole expansion of linearized} $f(R)$ \uppercase{gravity}\label{Sec:Stationary}}

We will focus on the multipole expansion of stationary fields in linearized $f(R)$ gravity with the irreducible Cartesian tensors
in this section. The word ``stationary" has two
meanings.  The first is that the sources do not depend on time $t'$, namely,
\begin{align}
\label{equ5.1}T^{\mu\nu}(t',\boldsymbol{x}')=T^{\mu\nu}(\boldsymbol{x}').%\tag{5.1}
\end{align}
The second is that the geometrical quantities, such as $\tilde h^{\mu\nu}$ and $R$, are independent of time.
%This condition is the basis of solving the stationary multipole expansion.

\subsection{The multipole expansion of stationary $\tilde{h}^{\mu\nu}$}

It is easy to obtain the multipole expansion of  stationary $\tilde{h}^{\mu\nu}$ from (\ref{equ4.2})---(\ref{equ4.4}).
In Ref.~\cite{Damour:1990gj}, $\overline{T}^{\mu\nu}_{l}$ in (\ref{equ4.4}) is written in a series form,
\begin{align}
\label{equ5.2}\overline{T}^{\mu\nu}_{l}(u,\boldsymbol{x}')=\sum_{k=0}^{\infty}\frac{(2l+1)!!}{(2k)!!(2l+2k+1)!!}
\frac{r'^{2k}}{c^{2k}}\frac{\partial^{2k}}{\partial u^{2k}}T^{\mu\nu}(u,\boldsymbol{x}').%\tag{5.2}
\end{align}
In stationary cases, only the $k=0$ term remains: \newpage

\begin{align}
\label{equ5.3}\overline{T}^{\mu\nu}_{l}(u,\boldsymbol{x}')=T^{\mu\nu}(\boldsymbol{x}').%\tag{5.3}
\end{align}
Eqs.~(\ref{equ4.3}) and (\ref{equ5.3}) result in that the mass-type and current-type source multipole moments do not depend on time, namely,
\begin{equation}\label{equ5.4}
\left\{\begin{array}{l}
\displaystyle\hat{M}_{I_{l}}=\frac{1}{c^{2}}\int d^{3}x'\hat{X'}_{I_{l}}\big(T^{00}(\boldsymbol{x}')+T^{aa}(\boldsymbol{x}')\big),\\
\displaystyle\hat{S}_{I_{l}}=\frac{1}{c}\int d^{3}x'\epsilon_{ab<i_{1}}\hat{X'}_{|a|i_{2}\cdots i_{l}>}T^{0b}(\boldsymbol{x}'),\quad l\geq1.
\end{array}\right.%\tag{5.4}
\end{equation}
With the two kinds of source multipole moments, Eq. (\ref{equ4.2}) leads to
\begin{equation}\label{equ5.5}
\left\{\begin{array}{l}
\displaystyle\tilde{h}^{00}(\boldsymbol{x})= -\frac{4G}{c^{2}}\sum_{l=0}^{\infty}\frac{(-1)^{l}}{l!}\hat{M}_{I_{l}}\partial_{I_{l}}\Big(\frac{1}{r}\Big),\\
\displaystyle\tilde{h}^{0i}(\boldsymbol{x})=-\frac{4G}{c^{3}}\sum_{l=1}^{\infty}\frac{(-1)^{l}l}{(l+1)!}\epsilon_{iab}\hat{S}_{aI_{l-1}}\partial_{bI_{l-1}}\Big(\frac{1}{r}\Big),\\
\displaystyle\tilde{h}^{ij}(\boldsymbol{x})=0,
\end{array}\right.
\end{equation}
which has the same form as the multipole expansion of $h^{\mu\nu}$ in GR.

\subsection{The multipole expansion of stationary $R^{(1)}$}

In the previous section, $R^{(1)}$ is decomposed into $R^{(1)}_{1}$ and $R^{(1)}_{2}$.  For stationary cases, however, the decomposition is not needed. By (\ref{equ4.6}) and (\ref{equ5.1}), we have
\begin{widetext}
\begin{align}
\label{equ5.6}R^{(1)}(\boldsymbol{x})&=-\frac{m^2\kappa}{4\pi}\int d^{3}x'\frac{T(\boldsymbol{x}')}{|\boldsymbol{x}-\boldsymbol{x}'|}
+\frac{m^2\kappa}{4\pi}\int d^{3}x'\int_{-\infty}^{t-\frac{|\boldsymbol{x}-\boldsymbol{x}'|}{c}} dt'\frac{mJ_{1}\left(mc\sqrt{(t-t')^{2}
-\frac{|\boldsymbol{x}-\boldsymbol{x}'|^{2}}{c^2}}\right)}{\sqrt{(t-t')^{2}-\frac{|\boldsymbol{x}-\boldsymbol{x}'|^{2}}{c^2}}}
T(\boldsymbol{x}').%\tag{5.6}
\end{align}
According to the result in Appendix B, we have
\begin{align}
\label{equ5.7}\int_{-\infty}^{t-\frac{|\boldsymbol{x}-\boldsymbol{x}'|}{c}}\frac{mJ_{1}\left(mc\sqrt{(t-t')^{2}-\frac{|\boldsymbol{x}-\boldsymbol{x}'|^{2}}{c^2}}\right)}{\sqrt{(t-t')^{2}-\frac{|\boldsymbol{x}-\boldsymbol{x}'|^{2}}{c^2}}} dt'
=\frac{1-\text{e}^{-m|\boldsymbol{x}-\boldsymbol{x}'|}}{|\boldsymbol{x}-\boldsymbol{x}'|},%\tag{5.7}
\end{align}
\end{widetext}
and then we obtain
\begin{align}
\label{equ5.8}R^{(1)}(\boldsymbol{x})&=-\frac{m^2\kappa}{4\pi}\int\frac{\text{e}^{-m|\boldsymbol{x}-\boldsymbol{x}'|}}{|\boldsymbol{x}-\boldsymbol{x}'|}
T(\boldsymbol{x}')d^{3}x'. %\tag{5.8}
\end{align}
The factor $|\boldsymbol{x}-\boldsymbol{x}'|^{-1}\text{e}^{-m|\boldsymbol{x}-\boldsymbol{x}'|}$ in the integrand in (\ref{equ5.8}) is the Yukawa potential.
Eq.~(\ref{equ3.63}) shows that $R^{(1)}$ satisfies the massive KG equation with an external source.
For stationary cases, this equation reduces to the screened Poisson equation:
\begin{align}
\label{equ5.9}\nabla^2 R^{(1)}-m^{2}R^{(1)}=m^{2}\kappa T^{(1)},%\tag{5.9}
\end{align}
where $\nabla^2$ is the Laplace operator in a flat space. Eq.~(\ref{equ5.8}) implies that the Green's function of
this differential equation is\pagebreak
\begin{align}
\label{equ5.10}\mathcal{G}(\boldsymbol{x};\boldsymbol{x}')=\frac{\text{e}^{-m|\boldsymbol{x}-\boldsymbol{x}'|}}{4\pi|\boldsymbol{x}-\boldsymbol{x}'|},%\tag{5.10}
\end{align}
and it satisfies
\begin{align}
\label{equ5.11}(\nabla^2-m^{2})\mathcal{G}(\boldsymbol{x};\boldsymbol{x}')=-\delta^{3}(\boldsymbol{x}-\boldsymbol{x}').%\tag{5.11}
\end{align}
Then, (\ref{equ5.8}) can be rewritten as
\begin{align}
\label{equ5.12}R^{(1)}(\boldsymbol{x})&=\int\mathcal{G}(\boldsymbol{x};\boldsymbol{x}')
\big(-m^2\kappa T^{(1)}(\boldsymbol{x}')\big)d^{3}x'.%\tag{5.12}
\end{align}

In order to derive the multipole expansion of $R^{(1)}(\boldsymbol{x})$, we need to deal with the Green's function (\ref{equ5.10}).
With the help of the differential equation (\ref{equ5.11}), the Green's function $\mathcal{G}(\boldsymbol{x};\boldsymbol{x}')$
can be written as
\begin{widetext}
\begin{align}
\label{equ5.13}\mathcal{G}(\boldsymbol{x};\boldsymbol{x}')&=\sum_{l=0}^{\infty}\frac{(2l+1)!!}{4\pi l!}mi_{l}(mr_<)k_{l}(mr_> )\hat{N}_{I_{l}}(\theta',\varphi')\hat{N}_{I_{l}}(\theta,\varphi)%\tag{5.13}
\end{align}
as shown in Appendix C, where
the meanings of $(\theta',\varphi')$, $(\theta,\varphi)$, $r_<$, and $r_>$ are the same as before,
\begin{align}
\label{equ5.14}i_{l}(z)&:=\sqrt{\frac{\pi}{2z}}I_{l+\frac{1}{2}}(z),\qquad
k_{l}(z):=\sqrt{\frac{2}{\pi z}}K_{l+\frac{1}{2}}(z)%\tag{5.14}.
\end{align}
are the spherical modified Bessel functions of $l$-order~\cite{Arfken1985}, and
$I_{l+1/2}(z)$, $K_{l+1/2}(z)$ are the modified Bessel functions of $(l+1/2)$-order.
Therefore, from (\ref{equ5.12}), we have
\begin{align}
\label{equ5.15}R^{(1)}(\boldsymbol{x})&=-\frac{m^3\kappa}{4\pi}\sum_{l=0}^{\infty}\frac{(-1)^l}{l!}
\int d^{3}x'(2l+1)!!i_{l}(mr_<)\hat{N}_{I_{l}}(\theta',\varphi')T^{(1)}(\boldsymbol{x}')(-1)^lk_{l}(mr_> )\hat{N}_{I_{l}}(\theta,\varphi)
.%\tag{5.15}
\end{align}
For the spherical modified Bessel functions, there are~\cite{Arfken1985},
\begin{align}
\label{equ5.16}i_{l}(z)&=z^l\Big(\frac{d}{zdz}\Big)^{l}\Big(\frac{\sinh{z}}{z}\Big),\qquad
k_{l}(z)=\frac{\text{e}^{-z}}{z}\sum_{k=0}^{l}\frac{(l+k)!}{k!(l-k)!}\frac{1}{(2z)^{k}}.%\tag{5.16}
\end{align}
Moreover, %we have
$$\text{e}^{-z}=(-1)^{l-k}\frac{d^{l-k}}{dz^{l-k}}\text{e}^{-z}.$$
With above three formulas, Eq. (\ref{equ5.15}) outside the source region (namely $r=r_>$ and $r'=r_<$) becomes
\begin{align}
\label{equ5.17}R^{(1)}(\boldsymbol{x})&\phantom{:}=-\frac{m^2\kappa}{4\pi}\sum_{l=0}^{\infty}\frac{(-1)^l}{l!}
\hat{Q}_{I_{l}}\hat{N}_{I_{l}}(\theta,\varphi)\sum_{k=0}^{l}\frac{(l+k)!}{(-2)^{k}k!(l-k)!}\frac{1}{r^{k+1}}\frac{d^{l-k}}{dr^{l-k}}\text{e}^{-mr},\\
%\tag{5.17a}\\
\label{equ5.18}\hat{Q}_{I_{l}}&:=\frac{(2l+1)!!}{m^{2l}}\int r'^l\Big(\frac{d}{r'dr'}\Big)^{l}\Big(\frac{\sinh{(mr')}}{mr'}\Big)\hat{N}_{I_{l}}(\theta',\varphi')T^{(1)}(\boldsymbol{x}')d^{3}x',%\tag{5.17b}
\end{align}
\end{widetext}
where $\hat{Q}_{I_{l}}$ is the stationary $l$-pole moment.

Now we make use of the STF technique to simplify (\ref{equ5.17}) and (\ref{equ5.18}).  For stationary cases,
Eq.~(\ref{equ2.14}) with $\epsilon=-1$ reduces to
\begin{align}
\label{equ5.19}\hat{\partial}_{I_{l}}&\Big(\frac{F(r)}{r}\Big)=\hat{N}_{I_{l}}\sum_{j=0}^{l}\frac{(l+j)!}{(-2)^{j}j!(l-j)!}\frac{F^{(l-j)}(r)}{r^{j+1}}.
%tag{5.18}
\end{align}
Inserting it into (\ref{equ5.17}) results in
\begin{align}
\label{equ5.20}R^{(1)}(\boldsymbol{x})
&=-\frac{m^{2}\kappa}{4\pi}\sum_{l=0}^{\infty}\frac{(-1)^l}{l!}\hat{Q}_{I_{l}}\hat\partial_{I_{l}}\Big(\frac{\text{e}^{-mr}}{r}\Big).%\tag{5.19a}
\end{align}
Because of (\ref{equ4.36}), Eq. (\ref{equ5.18}) can be written as
\begin{align}
\label{equ5.21}\hat{Q}_{I_{l}}&=\frac{(2l+1)!!}{m^{2l}}\int \hat{X'}_{I_{l}}\Big(\frac{d}{r'dr'}\Big)^{l}\Big(\frac{\sinh{(mr')}}{mr'}\Big)T^{(1)}(\boldsymbol{x}')d^{3}x'.%\tag{5.19b}
\end{align}

It is interesting to compare the above result with the multipole expansion of the Coulomb potential. For convenience,
we use $V$ to denote
a stationary massive scalar field, which satisfies
\begin{align}
\label{equ5.22}\nabla^2V-m^{2}V=-\frac{\rho}{\varepsilon_{0}},%\tag{5.20}
\end{align}
where $\rho(\boldsymbol{x})$ is the charge density, and $\varepsilon_{0}$ is a constant.
By (\ref{equ5.8}), we know that the solution of (\ref{equ5.22})  is
\begin{align}
\label{equ5.23}V(\boldsymbol{x})&=\frac{1}{4\pi\varepsilon_{0}}\int\frac{\text{e}^{-m|\boldsymbol{x}-\boldsymbol{x}'|}}{|\boldsymbol{x}-\boldsymbol{x}'|}
\rho(\boldsymbol{x}')d^{3}x',%\tag{5.21}
\end{align}
where the Green's function is still (\ref{equ5.10}).  With the help of (\ref{equ5.20}) and (\ref{equ5.21}), we easily get
the multipole expansion of $V$, namely,
\begin{align}
\label{equ5.24}V(\boldsymbol{x})
&=\frac{1}{4\pi\varepsilon_{0}}\sum_{l=0}^{\infty}\frac{(-1)^l}{l!}\hat{Q}_{I_{l}}\partial_{I_{l}}\Big(\frac{\text{e}^{-mr}}{r}\Big),%\tag{5.22a}
\end{align}
\begin{align}
\label{equ5.25}\hat{Q}_{I_{l}}&=\frac{(2l+1)!!}{m^{2l}}\int \hat{X'}_{I_{l}}\Big(\frac{d}{r'dr'}\Big)^{l}\Big(\frac{\sinh{(mr')}}{mr'}\Big)\rho(\boldsymbol{x}')d^{3}x'.%\tag{5.22b}
\end{align}

When $m=0$, Eq.~(\ref{equ5.22}) reduces to the Poisson equation for the Coulomb potential $V_{C}$,
\begin{align}
\label{equ5.26}\nabla^2V_{C}=-\frac{\rho}{\varepsilon_{0}},%\tag{5.23}
\end{align}
whose solution is %and then we know its expression by (\ref{equ5.21}), namely
\begin{align}
\label{equ5.27}V_{C}(\boldsymbol{x})&=\frac{1}{4\pi\varepsilon_{0}}\int\frac{1}{|\boldsymbol{x}-\boldsymbol{x}'|}
\rho(\boldsymbol{x}')d^{3}x'.%\tag{5.24}
\end{align}
%Again by (\ref{equ5.10}), we know
It is well known that the Green's function of (\ref{equ5.26}) is
\begin{equation}
\label{equ5.28}\mathcal{G}_{C}(\boldsymbol{x};\boldsymbol{x}')=\frac{1}{4\pi|\boldsymbol{x}-\boldsymbol{x}'|}.%\tag{5.25}
\end{equation}
Then, (\ref{equ5.27}) can be rewritten as %which satisfies
\begin{align}
\label{equ5.29}V_{C}(\boldsymbol{x})&=\int\mathcal{G}_{C}(\boldsymbol{x};\boldsymbol{x}')
\frac{\rho(\boldsymbol{x}')}{\varepsilon_{0}}d^{3}x'.%\tag{5.26}
\end{align}
%\clearpage
It is remarkable that the multipole expansion of $V_{C}$ cannot be obtained directly by setting $m=0$ in (\ref{equ5.24}) and (\ref{equ5.25}),
% to obtain the multipole expansion of $V_{C}$,
because (\ref{equ5.24}) and (\ref{equ5.25}) are not well defined at $m=0$.
However, the spherical modified Bessel functions
have the properties~\cite{Arfken1985}
\begin{align}
\label{equ5.30}i_{l}(z)&\approx\frac{z^l}{(2l+1)!!},\quad z\ll1,\\%\tag{5.27a}\\
\label{equ5.31}k_{l}(z)&\approx\frac{(2l-1)!!}{z^{l+1}},\quad z\ll1.%\tag{5.27b}
\end{align}
Substituting them in (\ref{equ5.13}) and then setting $m=0$, we obtain
\begin{align}
\label{equ5.32}\mathcal{G}_{C}(\boldsymbol{x};\boldsymbol{x}')&=\sum_{l=0}^{\infty}\frac{(2l-1)!!}{4\pi l!}\frac{(r_< )^l}{(r_> )^{l+1}}\hat{N}_{I_{l}}(\theta',\varphi')\hat{N}_{I_{l}}(\theta,\varphi), %\tag{5.28}
\end{align}
and thus %by (\ref{equ5.26}), we obtain
\begin{widetext}
\begin{align}
\label{equ5.33}V_{C}(\boldsymbol{x})&=\frac{1}{4\pi\varepsilon_{0}}\sum_{l=0}^{\infty}\frac{(-1)^l}{l!}\int r_<^l\hat{N}_{I_{l}}(\theta',\varphi')\rho(\boldsymbol{x}')d^{3}x'
\frac{(-1)^l(2l-1)!!}{r_> ^{l+1}}\hat{N}_{I_{l}}(\theta,\varphi).%\tag{5.29}
\end{align}
Outside the source region (namely $r=r_>$ and $r'=r_<$), we have
\begin{align}
\label{equ5.34}V_{C}(\boldsymbol{x})&\phantom{:}=\frac{1}{4\pi\varepsilon_{0}}\sum_{l=0}^{\infty}\frac{(-1)^l}{l!}
\hat{Q}_{CI_{l}}\hat{N}_{I_{l}}(\theta,\varphi)\frac{(-1)^l(2l-1)!!}{r^{l+1}},\\%\tag{5.30a}\\
\label{equ5.35}\hat{Q}_{CI_{l}}&:=\int r'^l\hat{N}_{I_{l}}(\theta',\varphi')\rho(\boldsymbol{x}')d^{3}x',%\tag{5.30b}
\end{align}
\end{widetext}
where $\hat{Q}_{CI_{l}}$ is the stationary $l$-pole moment.

Similarly, we can make use of the STF technique to simplify (\ref{equ5.34}) and (\ref{equ5.35}). When $F(r)=1$, (\ref{equ5.19}) reduces to
\begin{align}
\label{equ5.36}\hat{\partial}_{I_{l}}&\Big(\frac{1}{r}\Big)=\hat{N}_{I_{l}}(\theta,\varphi)\frac{(-1)^l(2l-1)!!}{r^{l+1}}.%\tag{5.31}
\end{align}
Then, (\ref{equ5.34}) and (\ref{equ5.35}) read
\begin{align}
\label{equ5.37}V_{C}(\boldsymbol{x})
&=\frac{1}{4\pi\varepsilon_{0}}\sum_{l=0}^{\infty}\frac{(-1)^l}{l!}\hat{Q}_{CI_{l}}\partial_{I_{l}}\Big(\frac{1}{r}\Big),\\%\tag{5.32a}
\label{equ5.38}\hat{Q}_{CI_{l}}&=\int \hat{X'}_{I_{l}}\rho(\boldsymbol{x}')d^{3}x',%\tag{5.32b}
\end{align}
respectively.

In (\ref{equ5.37}) and (\ref{equ5.38}), the 0th-order term of the Coulomb potential is
\begin{align}
\label{equ5.39}V_{C0}(\boldsymbol{x})
&=\frac{1}{4\pi\varepsilon_{0}}\frac{\hat{Q}_{C0}}{r},%\tag{5.33a}
\end{align}
where
\begin{align}
\label{equ5.40}\hat{Q}_{C0}&=\int \rho(\boldsymbol{x}')d^{3}x'%\tag{5.40}
\end{align}
is the total charge of the source. This shows that the 0th-order term of the Coulomb potential is equivalent to the potential of a point
charge whose charge is the total charge of the source, and which is located at the coordinate origin.

On the contrary, this conclusion does not hold for the Yukawa potential. This is because in (\ref{equ5.24}) and
(\ref{equ5.25}) the 0th-order term of the Yukawa potential is
\begin{align}
\label{equ5.41}V_{0}(\boldsymbol{x})
&=\frac{\hat{Q}_{0}}{4\pi\varepsilon_{0}}\frac{\text{e}^{-mr}}{r},%\tag{5.41}
\end{align}
where
\begin{align}
\label{equ5.42}\hat{Q}_{0}&=\int\frac{\sinh{(mr')}}{mr'} \rho(\boldsymbol{x}')d^{3}x'%\tag{5.42}
\end{align}
is not the total charge of the source.  It is obviously different from the above case of the Coulomb potential.

%%%%%%%%%%%%%%%%%%%%%%%%%%%%%%%%%%%%%%%%%%%%%%%%%%%%%%%
%%%%%%%%%%%%%%%%%%%%%%%%%%%%%%%%%%%%%%%%%%%%%%%%%%%%%%%
%%%           Stationary Cases                     %%%%
%%%%%%%%%%%%%%%%%%%%%%%%%%%%%%%%%%%%%%%%%%%%%%%%%%%%%%%
%%%%%%%%%%%%%%%%%%%%%%%%%%%%%%%%%%%%%%%%%%%%%%%%%%%%%%%

\subsection{The multipole expansion of stationary $h^{\mu\nu}$}
Once the multipole expansions of stationary $\tilde{h}^{\mu\nu}$ and $R^{(1)}$ are obtained,
the multipole expansion of $h^{\mu\nu}(\boldsymbol{x})$ is easily achieved by (\ref{equ4.1}). Firstly by
(\ref{equ3.64}), (\ref{equ5.20}), and $\kappa=8\pi G/c^4$, we get the multipole expansion of $-2a\eta^{\mu\nu}R^{(1)}(\boldsymbol{x})$,
and then we can derive the multipole expansion of $h^{\mu\nu}(\boldsymbol{x})$ by (\ref{equ5.5}), namely
%\begin{widetext}
\begin{equation}\label{equ5.43}
\left\{\begin{array}{l}
\displaystyle h^{00}(\boldsymbol{x})=-\frac{4G}{c^{2}}\sum_{l=0}^{\infty}\frac{(-1)^{l}}{l!}\hat{M}_{I_{l}}\partial_{I_{l}}\Big(\frac{1}{r}\Big)\\
\displaystyle \phantom{h^{00}(\boldsymbol{x})=}-\frac{2G}{3c^4}\sum_{l=0}^{\infty}\frac{(-1)^l}{l!}\hat{Q}_{I_{l}}\partial_{I_{l}}\Big(\frac{\text{e}^{-mr}}{r}\Big),\\
\displaystyle h^{0i}(\boldsymbol{x})=-\frac{4G}{c^{3}}\sum_{l=1}^{\infty}\frac{(-1)^{l}l}{(l+1)!}\epsilon_{iab}\hat{S}_{aI_{l-1}}\partial_{bI_{l-1}}\Big(\frac{1}{r}\Big),\\
\displaystyle h^{ij}(\boldsymbol{x})=
\frac{2G}{3c^4}\delta^{ij}\sum_{l=0}^{\infty}\frac{(-1)^l}{l!}\hat{Q}_{I_{l}}\partial_{I_{l}}\Big(\frac{\text{e}^{-mr}}{r}\Big).
\end{array}\right.%\tag{5.35}
\end{equation}
%\end{widetext}

%%%%%%%%%%%%%%%%%%%%%%%%%%%%%%%%%%%%%%%%%%%%%%%%%%%%%%%
%%%%%%%%%%%%%%%%%%%%%%%%%%%%%%%%%%%%%%%%%%%%%%%%%%%%%%%
%%%              Concluding Remarks                %%%%
%%%%%%%%%%%%%%%%%%%%%%%%%%%%%%%%%%%%%%%%%%%%%%%%%%%%%%%
%%%%%%%%%%%%%%%%%%%%%%%%%%%%%%%%%%%%%%%%%%%%%%%%%%%%%%%

\section{Conclusions and discussions \label{Sec:Conclusion}}

It has been shown in this paper that, similar to GR in Refs.~\cite{Thorne:1980ru,Blanchet:2013haa},
the field equations of $f(R)$ gravity can also be rewritten in the form of obvious wave equations in a fictitious
flat spacetime under the de Donder condition, even when the effective gravitational field amplitude $\tilde h^{\mu\nu}$ is not a perturbation.  The source
of the wave equation is the stress-energy pseudotensor of the matter fields and the
gravitational field. For the linearized $f(R)$ gravity, the corresponding field equations and
the effective stress-energy tensor of GWs are the same as the previous results in Ref.~\cite{Berry:2011pb}.
With this new form of field equations, some analytic approximations like
the post-Minkowskian method, the post-Newtonian method, the far zone expansion, and the perturbation in the small
mass limit, etc, can be applied into $f(R)$ gravity like GR, so that its nonlinearized effects
can be investigated more conveniently~\cite{Blanchet:2013haa}.

It has also been shown in the paper that the method of the multipole expansion with irreducible Cartesian tensors,
developed by Thorne, Blanchet, Damour, and Iyer~\cite{Thorne:1980ru,Damour:1990gj,Blanchet:1985sp},
can be applied in the linearized
$f(R)$ gravity.  Unlike GR, the gravitational field amplitude $h^{\mu\nu}$ contains two parts: one is the
tensor part associated with the effective gravitational field amplitude $\tilde h^{\mu\nu}$, and the other
is the scalar part associated with the linear part of Ricci scalar $R^{(1)}$, as in Refs.~\cite{Liang:2017ahj,Naf:2011za}.  The multipole expansion of the tensor
part $\tilde h^{\mu\nu}$ is the same as that of $h^{\mu\nu}$ in linearized GR because they satisfy the same equations
and are dealt with under the same gauge conditions.  The scalar part $R^{(1)}$ satisfies a massive KG equation with
an external source.  It contributes the multipole expansion nontrivially.  In this paper, we have
successfully derived the multipole expansion of $R^{(1)}$ in terms of irreducible Cartesian tensors.
Although the derivation process is somewhat tedious and complicated, the final expressions are simple enough and explicit.

With the help of the multipole expansions of $\tilde h^{\mu\nu}$ and $R^{(1)}$, the  multipole expansion of the
gravitational field amplitude $h^{\mu\nu}$ is obtained.  It shows that there exists monopole and dipole radiation
for $f(R)$ gravity in addition to the quadrupole and higher order moments, as pointed out in literature (see, for example, Ref~\cite{Naf:2011za}). Although the above conclusion is drawn in Ref~\cite{Naf:2011za}, but the slow-motion
approximation is adopted in addition to the weak-field approximation, and its method is also not built upon the STF
formalism, which is different from ours. Moreover, the moments of $R^{(1)}$ are only taken into
account up to hexadecapole moments in Ref~\cite{Naf:2011za}, but all the multipoles of $R^{(1)}$ are derived in the present paper.

%It should be remarked that the multipole expansions for massive scalar field and tensor field in terms of
%irreducible Cartesian tensors are actually defined by the Legendre polynomials or the scalar spherical harmonics functions.
%It is different from the multipole expansions for massive scalar field and vector field discussed in
%Ref.~\cite{Krause:1994ar}, which is defined by the Fourier series.\\
%{\red Remark on these two sentences.\\
%It seems that there is little logical connection between the former two sentences and these two sentences.  Maybe, we have to
%say more about \cite{Krause:1994ar} and then followed by these two sentences.}

As a particular case, the multipole expansion of the stationary fields in the linearized $f(R)$ gravity has also
been derived. The multipole expansion of the tensor part, namely the stationary $\tilde h^{\mu\nu}$, is easily obtained
from its time-dependent correspondence.  For the scalar part, namely the stationary scalar field $R^{(1)}$, the differential
equation becomes the screened Poisson equation. Its Green's function is the Yukawa potential.  The multipole expansions
of the Yukawa potential and thus the stationary massive scalar field are presented in an explicit form.

It is well known that when the mass parameter for the massive scalar field tends to zero, the stationary differential
equation reduces from the screened Poisson equation to the Poisson equation, and the Green's function reduces
from the Yukawa potential to Coulomb potential.  Although the multipole expansion of the Coulomb potential can not be obtained from that of the Yukawa potential automatically, it can be still derived by the asymptotical properties of
the spherical modified Bessel functions.
It should be remarked that the 0th-order term of the Yukawa potential, unlike that of the Coulomb potential,
is not equivalent to the potential of a point charge at the coordinate origin, whose charge is the total charge
of the source.

The STF formalism is an important method of multipole analysis with irreducible Cartesian tensors, and its usefulness has been emphasized and confirmed in GR~\cite{Thorne:1980ru,Blanchet:1985sp}.
%It has the advantage of rendering quite transparent the algebraic structure of vectorial and tensorial harmonicsexpansions~\cite{Damour:1990gj}.
%In this paper, the STF formalism has been applied into the multipole analysis of $f(R)$ gravity preliminarily like that in GR, we hope that the STF formalism will contribute to building the wave generation formalism of $f(R)$ gravity in the future.
The GW generation formalism of $f(R)$ gravity based on the STF technique
is worthy of being further investigated.

\acknowledgments{This work is supported by the National Natural Science Foundation of China (Grants No.~11690022) and by the Strategic Priority Research Program of the Chinese Academy of Sciences "Multi-waveband Gravitational Wave
Universe" (Grant No. XDB23040000).}
\appendix\label{appendix}
%%%%%%%%%%%%%%%%%%%%%%%%%%%%%%%%%%%%%%%%%%%%%%%%%%%%%%%
%%%%%%%%%%%%%%%%%%%%%%%%%%%%%%%%%%%%%%%%%%%%%%%%%%%%%%%
%%%               Appendix A                       %%%%
%%%%%%%%%%%%%%%%%%%%%%%%%%%%%%%%%%%%%%%%%%%%%%%%%%%%%%%
%%%%%%%%%%%%%%%%%%%%%%%%%%%%%%%%%%%%%%%%%%%%%%%%%%%%%%%
\section{DERIVATION OF (\ref{equ4.53})}
(\ref{equ4.52}) and (\ref{equ4.49}) can be simplified as
\begin{align}
\label{equA1}\qquad c_{l}&=\frac{2l+1}{2}\int_{-1}^{1}\mathcal{K}(x)P_{l}(x)dx,\\%\tag{A.1}\\
\label{equA2}\mathcal{K}(x)&=\frac{J_{1}\left(\sqrt{\tilde{b}(x-\nu_{2})}\right)}{\sqrt{\tilde{b}(x-\nu_{2})}},\quad x\geq\nu_{2},%\tag{A.2}
\end{align}
where
\begin{align}
\label{equA3}x:=\cos{\tilde{\theta}},\qquad
\tilde{b}:=2m^2rr'.%\tag{A.3}
\end{align}
Replacing the integral variable $x$ with $$y:=\sqrt{\tilde{b}(x-\nu_{2})}$$ in (\ref{equA1}) gives
\begin{align}
\label{equA4}&c_{l}=\frac{2l+1}{\tilde{b}}\int_{\sqrt{-\tilde{b}(1+\nu_{2})}}^{\sqrt{\tilde{b}(1-\nu_{2})}}J_{1}(y)P_{l}\big(\frac{y^2}{\tilde{b}}+\nu_{2}\big)dy.
%\tag{A.4}
\end{align}
By the definition of the Legendre polynomial and the binomial expansion, we get
\begin{widetext}
\begin{align}
\label{equA5}&c_{l}=\frac{2l+1}{\tilde{b}}\sum_{k=0}^{[\frac{l}{2}]}\frac{(-1)^k}{2^lk!}\frac{(2l-2k)!}{(l-k)!(l-2k)!}
\sum_{j=0}^{l-2k}\frac{(l-2k)!}{j!(l-2k-j)!}\frac{\nu_{2}^{l-2k-j}}{\tilde{b}^{j}}
\int_{\sqrt{-\tilde{b}(1+\nu_{2})}}^{\sqrt{\tilde{b}(1-\nu_{2})}}y^{2j}J_{1}(y)dy.%\tag{A.5}
\end{align}
Further, from the series representation of the Bessel function and the definition of the generalized hypergeometric
function~\cite{Brychkov2008}
\begin{align}
\label{equA6}&{}_{p}F_{q}(a_{1},\cdots,a_{p};b_{1},\cdots,b_{q};z):=\sum_{k=0}^{\infty}\frac{(a_{1})_{k}\cdots (a_{p})_{k}}{(b_{1})_{k}\cdots(b_{q})_{k}}\frac{z^{k}}{k!},%\tag{A.8}
\end{align}
where
\begin{align}
\label{equA7}&(a_{i})_{k}:=a_{i}(a_{i}+1)\cdots(a_{i}+k-1)%\tag{A.9}
\end{align}
is the Pochhammer symbol,
it is easy to show
\begin{align}
\label{equA8}&\int y^{2j}J_{1}(y)dy=\frac{y^{2j+2}}{4(j+1)}\;{}_{1}F_{2}\Big(j+1;2,j+2;-\frac{y^2}{4}\Big)+C,%\tag{A.6}
\end{align}
where $C$ is a constant of integration. It leads to
\begin{align}
\int_{\sqrt{-\tilde{b}(1+\nu_{2})}}^{\sqrt{\tilde{b}(1-\nu_{2})}}y^{2j}J_{1}(y)dy=&\frac{\tilde{b}^{j+1}(1-\nu_{2})^{j+1}}{4(j+1)}
\;{}_{1}F_{2}\Big(j+1;2,j+2;-\frac{\tilde{b}(1-\nu_{2})}{4}\Big)\notag\\
\label{equA9}&-\frac{\tilde{b}^{j+1}(-1-\nu_{2})^{j+1}}{4(j+1)}\;{}_{1}F_{2}\Big(j+1;2,j+2;\frac{\tilde{b}(1+\nu_{2})}{4}\Big).%\tag{A.7}
\end{align}
Inserting (\ref{equA9}) into (\ref{equA5}) gives
\begin{align}
\label{equA10}c_{l}=&\frac{2l+1}{\tilde{b}}\sum_{k=0}^{[\frac{l}{2}]}\frac{(-1)^k}{2^lk!}\frac{(2l-2k)!}{(l-k)!(l-2k)!}(A_{lk}-B_{lk}),%\tag{A.10}
\end{align}
where
\begin{align}
\label{equA11}A_{lk}&:=\tilde{b}\sum_{j=0}^{l-2k}\frac{(l-2k)!}{j!(l-2k-j)!}
\frac{\nu_{2}^{l-2k-j}(1-\nu_{2})^{j+1}}{4(j+1)}\;{}_{1}F_{2}\Big(j+1;2,j+2;-\frac{\tilde{b}(1-\nu_{2})}{4}\Big),\\%\tag{A11a}\\
\label{equA12}B_{lk}&:=\tilde{b}\sum_{j=0}^{l-2k}\frac{(l-2k)!}{j!(l-2k-j)!}
\frac{\nu_{2}^{l-2k-j}(-1-\nu_{2})^{j+1}}{4(j+1)}\;{}_{1}F_{2}\Big(j+1;2,j+2;\frac{\tilde{b}(1+\nu_{2})}{4}\Big).%\tag{A11b}
\end{align}

Now we simplify (\ref{equA11}) and (\ref{equA12}).  From (\ref{equA6}),
\begin{align}
\label{equA13}{}_{1}F_{2}\Big(j+1;2,j+2;-\frac{\tilde{b}z}{4}\Big)=
\sum_{\mu=0}^{\infty}\frac{j+1}{j+\mu+1}\frac{1}{\mu!(\mu+1)!}\Big(-\frac{\tilde{b}z}{4}\Big)^{\mu}.%\tag{A.12}
\end{align}
By use of (\ref{equA13}), Eqs.~(\ref{equA11}) and (\ref{equA12}) can be rewritten as
\begin{align}
\label{equA14}A_{lk}&=-\nu_{2}^{l-2k}\sum_{\mu=0}^{\infty}\Big(-\frac{\tilde{b}(1-\nu_{2})}{4}\Big)^{\mu+1}\frac{1}{(\mu+1)!(\mu+1)!}
\;{}_{2}F_{1}\Big(2k-l,\mu+1;\mu+2;\frac{\nu_{2}-1}{\nu_{2}}\Big),\\%\tag{A13a}\\
\label{equA15}B_{lk}&=-\nu_{2}^{l-2k}\sum_{\mu=0}^{\infty}\Big(-\frac{\tilde{b}(-1-\nu_{2})}{4}\Big)^{\mu+1}\frac{1}{(\mu+1)!(\mu+1)!}
\;{}_{2}F_{1}\Big(2k-l,\mu+1;\mu+2;\frac{\nu_{2}+1}{\nu_{2}}\Big),%\tag{A13b}
\end{align}
where $\,{}_{2}F_{1}$ is the Gauss hypergeometric function [also cf.~(\ref{equA6})].  With the help of the property of $\,{}_{2}F_{1}$ about
the Pfaff transformation~\cite{Abramowitz1965}, we obtain
\begin{align}
\label{equA16}{}_{2}F_{1}\Big(2k-l,\mu+1;\mu+2;\frac{\nu_{2}-1}{\nu_{2}}\Big)&=\left (\frac{1}{\nu_{2}}\right )^{l-2k}{}_{2}F_{1}\Big(2k-l,1;\mu+2;1-\nu_{2}\Big),\\%\tag{A14a}\\
\label{equA17}{}_{2}F_{1}\Big(2k-l,\mu+1;\mu+2;\frac{\nu_{2}+1}{\nu_{2}}\Big)&=\left (-\frac{1}{\nu_{2}}\right )^{l-2k}{}_{2}F_{1}\Big(2k-l,1;\mu+2;1+\nu_{2}\Big).%\tag{A14b}
\end{align}
It follows from (\ref{equA6}) that
 \begin{align}
\label{equA18}{}_{2}F_{1}\big(2k-l,1;\mu+2;z\big)=
\sum_{j=0}^{l-2k}(-1)^j\frac{(l-2k)!}{(l-2k-j)!}\frac{(\mu+1)!}{(\mu+j+1)!}z^j.%\tag{A15}
\end{align}
The substitution of the above three formulas into (\ref{equA14}) and (\ref{equA15}) gives
\begin{align}
\label{equA19}A_{lk}&=-\sum_{j=0}^{l-2k}\big(-(1-\nu_{2})\big)^{j}\frac{(l-2k)!}{j!(l-2k-j)!}
\Bigg({}_{0}F_{1}\Big(j+1;-\frac{\tilde{b}(1-\nu_{2})}{4}\Big)-1\Bigg),\\%\tag{A16a}\\
\label{equA20}B_{lk}&=-\sum_{j=0}^{l-2k}(-1)^{l-2k}(-1-\nu_{2})^{j}\frac{(l-2k)!}{j!(l-2k-j)!}
\left ({}_{0}F_{1}\Big(j+1;-\frac{\tilde{b}(-1-\nu_{2})}{4}\Big)-1\right ),%\tag{A16b}
\end{align}
where ${}_{0}F_{1}$ is also a generalized hypergeometric function.  Then,
\begin{align}
\label{equA21}A_{lk}-B_{lk}&=-\sum_{j=0}^{l-2k}\big(-(1-\nu_{2})\big)^{j}\frac{(l-2k)!}{j!(l-2k-j)!}
{}_{0}F_{1}\Big(j+1;-\frac{\tilde{b}(1-\nu_{2})}{4}\Big)\notag\\
+&(-1)^{l}\sum_{j=0}^{l-2k}(-1-\nu_{2})^{j}\frac{(l-2k)!}{j!(l-2k-j)!}
{}_{0}F_{1}\Big(j+1;-\frac{\tilde{b}(-1-\nu_{2})}{4}\Big).%\tag{A17}
\end{align}
By (\ref{equA6}), we know
\begin{align}
\label{equA22}(1-\nu_{2})^{j}{}_{0}F_{1}\Big(j+1;-\frac{\tilde{b}(1-\nu_{2})}{4}\Big)&=
\sum_{k=0}^{\infty}\frac{1}{k!}\Big(-\frac{\tilde{b}}{4}\Big)^{k}\frac{(1-\nu_{2})^{j+k}}{(j+1)_{k}},\\%\tag{A18a}\\
\label{equA23}(-1-\nu_{2})^{j}{}_{0}F_{1}\Big(j+1;-\frac{\tilde{b}(-1-\nu_{2})}{4}\Big)&=
\sum_{k=0}^{\infty}\frac{1}{k!}\Big(-\frac{\tilde{b}}{4}\Big)^{k}\frac{(-1-\nu_{2})^{j+k}}{(j+1)_{k}}.%\tag{A18b}
\end{align}
For an arbitrary function $\tilde{f}(\nu_{2})$, we define
\begin{align}
\label{equA24}\left(\frac{d}{d\nu_{2}}\right)_{\pm}^{-1}\tilde{f}(\nu_{2}):&=-\int_{\nu_{2}}^{\pm1}\tilde{f}(s)ds.
%,\qquad \left(\frac{d}{d\nu_{2}}\right)_{-}^{-1}\tilde{f}(\nu_{2}):=-\int_{\nu_{2}}^{-1}\tilde{f}(s)ds.\tag{A19}
\end{align}
In particular,
\begin{align}
\label{equA25}\left(\frac{d}{d\nu_{2}}\right)_{+ }^{-k}( 1-\nu_{2})^{j}&=(-1)^k\frac{(1-\nu_{2})^{j+k}}{(j+1)_{k}}
,\qquad\left(\frac{d}{d\nu_{2}}\right)_{-}^{-k}(-1-\nu_{2})^{j}=(-1)^k\frac{(-1-\nu_{2})^{j+k}}{(j+1)_{k}},%\tag{A20}
\end{align}
which lead to
\begin{align}
\label{equA26}(1-\nu_{2})^{j}\;{}_{0}F_{1}\Big(j+1;-\frac{\tilde{b}(1-\nu_{2})}{4}\Big)&=
\sum_{k=0}^{\infty}\frac{1}{k!}\Big(\frac{\tilde{b}}{4}\Big)^{k}\Big(\frac{d}{d\nu_{2}}\Big)_{+}^{-k}(1-\nu_{2})^{j}
=\exp{\left(\frac{\tilde{b}}{4}\left(\frac{d}{d\nu_{2}}\right)_{+}^{-1}\right)}(1-\nu_{2})^{j},\\%\tag{A21a}\\
\label{equA27}(-1-\nu_{2})^{j}\;{}_{0}F_{1}\Big(j+1;-\frac{\tilde{b}(-1-\nu_{2})}{4}\Big)&=
\sum_{k=0}^{\infty}\frac{1}{k!}\Big(\frac{\tilde{b}}{4}\Big)^{k}\Big(\frac{d}{d\nu_{2}}\Big)_{-}^{-k}(-1-\nu_{2})^{j}
=\exp{\left(\frac{\tilde{b}}{4}\left(\frac{d}{d\nu_{2}}\right)_{-}^{-1}\right)}(-1-\nu_{2})^{j}.%\tag{A21b}
\end{align}
With this two formulas, (\ref{equA21}) can be simplified to
\begin{align}
\label{equA28}A_{lk}-B_{lk}&=
\exp{\left(\frac{\tilde{b}}{4}\left(\frac{d}{d\nu_{2}}\right)_{-}^{-1}\right)}\nu_{2}^{l-2k}
-\exp{\left(\frac{\tilde{b}}{4}\left(\frac{d}{d\nu_{2}}\right)_{+}^{-1}\right)}\nu_{2}^{l-2k},%\tag{A22}
\end{align}
and then inserting it into (\ref{equA10}) gives
\begin{align}
\label{equA29}c_{l}=&\frac{2l+1}{\tilde{b}}\left[\exp{\left(\frac{\tilde{b}}{4}\left(\frac{d}{d\nu_{2}}\right)_{-}^{-1}\right)}
-\exp{\left(\frac{\tilde{b}}{4}\left(\frac{d}{d\nu_{2}}\right)_{+}^{-1}\right)}\right ]P_{l}(\nu_{2})%\tag{A23}
\end{align}
by the definition of the Legendre polynomial. By (\ref{equA3}), we know (\ref{equ4.53}) holds.
\end{widetext}
%%%%%%%%%%%%%%%%%%%%%%%%%%%%%%%%%%%%%%%%%%%%%%%%%%%%%%%
%%%%%%%%%%%%%%%%%%%%%%%%%%%%%%%%%%%%%%%%%%%%%%%%%%%%%%%
%%%               Appendix B                       %%%%
%%%%%%%%%%%%%%%%%%%%%%%%%%%%%%%%%%%%%%%%%%%%%%%%%%%%%%%
%%%%%%%%%%%%%%%%%%%%%%%%%%%%%%%%%%%%%%%%%%%%%%%%%%%%%%%
%\newpage
\section{DERIVATION OF (\ref{equ5.7})}
Define
\begin{align}
\label{equB1}\tilde{a}:=\frac{|\boldsymbol{x}-\boldsymbol{x}'|}{c}%\tag{B.1}
\end{align}
and
\begin{align}
\label{equB2}t'=t-\tilde{a}\cosh{\psi},\qquad \psi\in[0,+\infty).%\tag{B.1}
\end{align}
With the new variable $\psi$, the left-hand side of Eq.~(\ref{equ5.7}) becomes
\begin{align}
\label{equB3}&\int_{-\infty}^{t-\frac{|\boldsymbol{x}-\boldsymbol{x}'|}{c}}\frac{mJ_{1}\left(mc\sqrt{(t-t')^{2}-\frac{|\boldsymbol{x}-\boldsymbol{x}'|^{2}}{c^2}}\right)}{\sqrt{(t-t')^{2}-\frac{|\boldsymbol{x}-\boldsymbol{x}'|^{2}}{c^2}}} dt' \notag \\
%=\int_{-\infty}^{-\tilde{a}}\frac{mJ_{1}(mc\sqrt{\tau^2-\tilde{a}^2})}{\sqrt{\tau^2-\tilde{a}^2}}d\tau.%\tag{B.2}
%\end{align}
%Replacing the integral variable $\tau$ with $\theta$ in (\ref{equB.2}) by
%$$\tau:=-\tilde{a}\cosh{\theta}$$
%gives
%\begin{align}
%\label{equB.3}\int_{-\infty}^{-\tilde{a}}\frac{mJ_{1}(mc\sqrt{\tau^2-\tilde{a}^2})}{\sqrt{\tau^2-\tilde{a}^2}}d\tau
=&\int_{0}^{\infty}mJ_{1}(mc\tilde{a}\sinh{\psi})d\psi.%\tag{B.3}
\end{align}
By Ref.~\cite{Gradshteyn2007}, we know the formula
\begin{align}
\label{equB4}\int_{0}^{\infty}J_{\mu+\nu}(2z\sinh{t})\cosh\big((\mu-\nu)t\big)dt
=I_{(\mu}(z)K_{\nu)}(z),%\tag{B.4}
\end{align}
where $\mu$, $\nu$, and $z$ have to satisfy
\begin{align}
\label{equB5}\text{Re}(\mu+\nu)>-1,\
|\text{Re}(\mu-\nu)|<\frac{3}{2},\
z>0,%\tag{B5c}
\end{align}
and $I_{\mu}(z)$ and $K_{\mu}(z)$ are the modified Bessel functions of $\mu$-order. Applying (\ref{equB4})
into (\ref{equB3}) gives
\begin{align}
\label{equB6}\int_{0}^{\infty}mJ_{1}(mc\tilde{a}\sinh{\psi})d\psi
=mI_{\frac{1}{2}}\Big(\frac{mc\tilde{a}}{2}\Big)K_{\frac{1}{2}}\Big(\frac{mc\tilde{a}}{2}\Big).%\tag{B6}
\end{align}
Since %Again, we make use of the properties of $I_{\mu}(z)$ and $K_{\mu}(z)$
\begin{align}
\label{equB7}&I_{\frac{1}{2}}(z)=\sqrt{\frac{2}{\pi z}}\sinh{z},\
K_{\frac{1}{2}}(z)=\sqrt{\frac{\pi}{2z}}\text{e}^{-z},%\tag{B7b}
\end{align}
Eq.~(\ref{equB6}) reads%to obtain
\begin{align*}
\label{equB8}\int_{0}^{\infty}mJ_{1}(mc\tilde{a}\sinh{\psi})d\psi
=\frac{1-\text{e}^{-mc\tilde{a}}}{c\tilde{a}}.\tag{B8}
\end{align*}
Finally by (\ref{equB1}) and (\ref{equB3}), we know that (\ref{equ5.7}) holds.
%%%%%%%%%%%%%%%%%%%%%%%%%%%%%%%%%%%%%%%%%%%%%%%%%%%%%%%
%%%%%%%%%%%%%%%%%%%%%%%%%%%%%%%%%%%%%%%%%%%%%%%%%%%%%%%
%%%               Appendix C                       %%%%
%%%%%%%%%%%%%%%%%%%%%%%%%%%%%%%%%%%%%%%%%%%%%%%%%%%%%%%
%%%%%%%%%%%%%%%%%%%%%%%%%%%%%%%%%%%%%%%%%%%%%%%%%%%%%%%
\section{DERIVATION \uppercase{of} (\ref{equ5.13})}
In the spherical coordinates (\ref{equ2.2}), Eq.~(\ref{equ5.11}) has the form
\begin{align}
\label{equC1} &\frac{1}{r^2}\frac{\partial}{\partial r}\Big(r^2\frac{\partial\mathcal{G}}{\partial r}\Big)
-\frac{\hat{L}^2}{r^2}\mathcal{G}-m^{2}\mathcal{G} \notag \\
=&-\dfrac{\delta(r-r')\delta(\theta-\theta')\delta(\varphi-\varphi')}{r^2\sin{\theta}},%\tag{C1a}
\end{align}
where
\begin{align}
\label{equC2}\hat{L}^2=
-\frac{1}{\sin{\theta}}\frac{\partial}{\partial\theta}\Big(\sin{\theta}\frac{\partial}{\partial\theta}\Big)
-\frac{1}{\sin^2{\theta}}\frac{\partial^2}{\partial\varphi^2}%\tag{C1b}
\end{align}
satisfies
\begin{align}
\label{equC3}\hat{L}^2Y^{lm'}(\theta,\varphi)=l(l+1)Y^{lm'}(\theta,\varphi).%\tag{C.2}
\end{align}
We consider the following Ansatz
 ~\cite{Sizhu1987} for Eq.~(\ref{equC1}),
\begin{align}
\label{equC4}\mathcal{G}(\boldsymbol{x};\boldsymbol{x}')
=\sum_{l=0}^{\infty}\sum_{m'=-l}^{l}g^{}_{l}(r,r')Y^{lm'*}(\theta',\varphi')Y^{lm'}(\theta,\varphi). %\tag{C.3}
\end{align}
It makes Eq.~(\ref{equC1}) reduce to an ordinary differential equation,
\begin{align}
\label{equC5}\frac{d^2g^{}_{l}}{dr^2}+\frac{2}{r}\frac{dg^{}_{l}}{dr}
-\frac{l(l+1)}{r^2}g^{}_{l}-m^{2}g^{}_{l}
=-\frac{\delta(r-r')}{r^2}.%\tag{C.4}
\end{align}

Firstly, we should solve the homogeneous differential equation of (\ref{equC5}).
If we define
\begin{align}
\label{equC6}u^{}_{l}(r,r'):=\sqrt{r}g^{}_{l}(r,r'),%\tag{C.5}
\end{align}
the homogeneous differential equation of (\ref{equC5}) is equivalent to
\begin{align}
\label{equC7}r^2\frac{d^2u^{}_{l}}{dr^2}+r\frac{du^{}_{l}}{dr}
-\left (m^{2}r^2+\Big(l+\frac{1}{2}\Big)^2\right )u^{}_{l}=0,%\tag{C.6}
\end{align}
which is the modified Bessel differential equation, and
its general solution is
\begin{align}
\label{equC8}u^{}_{l}(r,r')=AI_{l+\frac{1}{2}}(mr)+BK_{l+\frac{1}{2}}(mr),%\tag{C.6}
\end{align}
where $A,B$ are two constants of integration.
By (\ref{equ5.14}), (\ref{equC6}) and (\ref{equC8}), we know that the general solution of the homogeneous differential equation of (\ref{equC5}) is
\begin{align}
\label{equC9}g^{(h)}_{l}(r,r')=C_{1}i_{l}(mr)+C_{2}k_{l}(mr),%\tag{C.7}
\end{align}
where $C_{1},C_{2}$ are two constants related to $A$ and $B$.

Next, we try to solve the general solution of (\ref{equC5}). According to Ref.~\cite{Sizhu1987},
the general solution of (\ref{equC5}) is
\begin{align}
\label{equC10}g^{}_{l}(r,r')=&g^{(h)}_{l}(r,r')-k_{l}(mr)\int_{0}^{r}\frac{\delta{(\varepsilon-r')}i_{l}(m\varepsilon)}{\varepsilon^2\Delta{(\varepsilon)}}
d\varepsilon\notag\\
&-i_{l}(mr)\int_{r}^{\infty}\frac{\delta{(\varepsilon-r')}k_{l}(m\varepsilon)}{\varepsilon^2\Delta{(\varepsilon)}}d\varepsilon,%\tag{C.8}
\end{align}
where
\begin{align}
\label{equC11}%\bigtriangleup
\Delta{(r)}:= \begin{vmatrix}
i_{l}(mr)&k_{l}(mr)\\
\dfrac{d}{dr}\big(i_{l}(mr)\big)&\dfrac{d}{dr}\big(k_{l}(mr)\big)
\end{vmatrix}%\tag{C.9}
\end{align}
is the Wronskian determinant of the homogeneous differential equation of (\ref{equC5}).
Again by Eq.~(8.140) in Ref.~\cite{Sizhu1987}, we derive
\begin{align}
\label{equC12}\Delta{(r)}=\exp{\Big(-\int\frac{2}{r}dr\Big)}=\frac{C}{r^2},
%\tag{C.10}
\end{align}
where $C$ is a constant of integration,
and then we obtain
\begin{align}
\label{equC13}\Delta{(r)}=\frac{r_{1}^2}{r^2}\Delta{(r_{1})},
%\tag{C.11}
\end{align}
where $r_{1}$ is an arbitrary constant.
The spherical modified Bessel functions have the following asymptotical behaviors for sufficient large $z$ \cite{Arfken1985}:
\begin{align}
\label{equC14}i_{l}(z)&\approx\frac{\text{e}^z}{2z},\\%\tag{C13a}\\
\label{equC15}k_{l}(z)&\approx\frac{\text{e}^{-z}}{z}.%\tag{C13b}
\end{align}
If $r_{1}$ is large enough, then by (\ref{equC11}), (\ref{equC14}), and (\ref{equC15}), we have
\begin{align}
\label{equC16}\Delta{(r_{1})}\approx-\frac{1}{mr_{1}^2},%\tag{C.13}
\end{align}
and finally by (\ref{equC13}), we get
\begin{align}
\label{equC17}\Delta{(r)}=-\frac{1}{mr^2}.%\tag{C.14}
\end{align}

Inserting (\ref{equC17}) into (\ref{equC10}) gives
\begin{align}
\label{equC18}g^{}_{l}(r,r')=g^{(h)}_{l}(r,r')+mi_{l}(mr_{<})k_{l}(mr_{>}).%\tag{C.15}
\end{align}
The boundary condition of (\ref{equC5}) is
\begin{align}
\label{equC19}|g^{}_{l}(r,r')|\ll\infty,\qquad r\rightarrow0,\\%\tag{C17a}\\
\label{equC20}g^{}_{l}(r,r')\rightarrow0,\qquad r\rightarrow\infty.%\tag{C17b}
\end{align}
Applying them to (\ref{equC18}), and by using (\ref{equ5.30}), (\ref{equ5.31}), (\ref{equC14}), and (\ref{equC15}), we
acquire
\begin{align}
\label{equC21}g^{}_{l}(r,r')=mi_{l}(mr_{<})k_{l}(mr_{>}).%\tag{C18}
\end{align}
Hence, (\ref{equ5.13}) is obviously valid after the insert of (\ref{equC21}) in (\ref{equC4}) and then use of (\ref{equ2.15}).
%\section*{References}
%\bibliographystyle{unsrt}
%\bibliography{mybbib}

\end{document}